\documentclass[referee]{raa}            

\usepackage{amssymb,amsmath}
\usepackage{amsfonts}
\usepackage{amssymb}
\usepackage[titletoc,title]{appendix}
\usepackage{booktabs}
\usepackage[mmddyyyy]{datetime}
\usepackage{dutchcal}
\usepackage{enumitem}
\usepackage{float}
\usepackage{graphicx,times}  
\usepackage[utf8]{inputenc}
\usepackage{lineno}
\usepackage{lscape}
\usepackage{longtable}
\usepackage{natbib}
\usepackage[percent]{overpic}
\usepackage{pgfplots}
\usepackage{relsize}
\usepackage{siunitx}
\usepackage{tabularx}
\usepackage{tikz}
 \usepackage{url,textcomp}

\modulolinenumbers[5]
\numberwithin{equation}{section}

\usetikzlibrary{calc}
\usetikzlibrary{plotmarks}
\usetikzlibrary{positioning}
\usetikzlibrary{fit}
\usetikzlibrary{decorations.pathmorphing}
\usetikzlibrary{backgrounds}
\makeatother
\modulolinenumbers[5]
\numberwithin{equation}{section}
\pgfplotsset{compat = newest}
\pgfplotsset{ legend style={font=\tiny} }
\definecolor{bgreen}{rgb}{0.0,0.5,0.0}
\definecolor{bblue}{rgb}{0.0,0.0,0.9}
\definecolor{bgold}{rgb}{0.7,0.5,0.0}
\definecolor{bred}{rgb}{0.9,0.0,0.0}
\bibpunct{(}{)}{;}{a}{}{,}

\begin{document}

   \title{An orbit-averaged generalized-Landau kinetic equation for the relaxation evolution of finite weakly-coupled star clusters: `Discreteness' stochastic acceleration and anti-normalization
}

   \volnopage{Vol.0 (20xx) No.0, 000--000}      
   \setcounter{page}{1}          

   \author{Yuta Ito
      \inst{1,2}
   }

   \institute{Department of Physics, The Graduate Center of CUNY/ 365 Fifth Avenue, New York, NY 10016, USA; {\it yutaito30@gmail.com}\\
        \and
             Department of Physics and Astronomy, The College of Staten Island of CUNY/ 2800 Victory Boulevard, Staten Island, NY 10314, USA\\
\vs\no
   {\small Received~~20xx month day; accepted~~20xx~~month day}}

\abstract{In the relaxation evolution of finite weakly-coupled  star clusters, stars undergo stochastic acceleration due to the `discreteness' of the clusters (the finiteness of the total stellar number), in addition to fundamental two-body relaxation processes. The acceleration is the essential non-collective many-body relaxation process. However, existing works have never detailed the `discreteness' stochastic acceleration and the corresponding mathematical model, i.e., the generalized-Landau (g-Landau) kinetic equation for the stellar distribution function. 
The present paper shows the kinetic formulation of an orbit-averaged g-Landau equation in action-angle coordinates, beginning with Bogoliubov–Born–Green–Kirkwood–Yvon hierarchy. We show that only the g-Landau equation can satisfy the anti-normalization condition among existing approximated star-cluster kinetic equations if gravitational polarization is neglected. Accordingly, only the equation can correctly define the total energy and the total number of stars in phase space. It also  holds the conservation laws and the \emph{H}-theorem.  We further show that stars undergoing the discreteness stochastic acceleration can hold these physical properties independently of stars experiencing the two-body relaxation. Lastly, we suggest that the stochastic acceleration tends to isotropize the stellar DF and to drive star clusters into a stationary state in the early relaxation-evolution stage before the two-body relaxation thermalizes stars.
}

   \authorrunning{Y. Ito}            
   \titlerunning{Orbit-Averaged generalized Landau equation}  

   \maketitle

%
%
\section{Introduction}\label{sec:intro}

We consider an ideal isolated star cluster of $N$-stars, i.e., $N$-particles of equal masses $m$ interacting with each other via the Newtonian pairwise force under no influence of the tidal effect from its host galaxy. The number $N$ of stars is approximately $10^{4}\sim 10^{6}$. The standard kinetic theory separates stars into `test' star and $(N-1)$-`field' stars.\footnote{The test star is considered a statistically ’typical’ star. Its motion is determined from the MF potential and Newtonian pairwise potential from the field stars (or background stars).
However, the difference between the test star and field stars is practically subtle at the end of the standard kinetic formulation since the formulation assumes that all the stars indistinguishable. Hence, the motion of the test star is considered typical for field stars.} The acceleration $\boldsymbol{a}_{1}$ of star 1 (or the test star) is only subject to the vector sum of pairwise-Newtonian potential forces due to stars 2, 3, $\cdots$, $N$ (or $(N-1)$-field stars);
\begin{align}
\boldsymbol{a}_{1}=\sum^{N}_{i=2}\boldsymbol{a}_{1i},
\end{align}
where the acceleration $\boldsymbol{a}_{1i}$ of the test star at position vector $\boldsymbol{r}_{1}$ is determined by the pairwise potential $\phi_{1i}$ from a field star $i$ at $\boldsymbol{r}_{i}$ as follows
\begin{align}
&\boldsymbol{a}_{1i}\equiv-\nabla_{1} \phi_{1i}=-\frac{\partial}{\partial \boldsymbol{r}_1}\left(\frac{-Gm}{\mid \boldsymbol{r}_{1}-\boldsymbol{r}_{i} \mid}\right),\qquad\qquad\quad (i=2,\cdots,N)
\end{align}
where $G$ is the Newton gravitational constant. However, direct $N$-body numerical simulations for the relaxation evolution of star clusters are numerically expensive ($\sim$ a few months through years of CPU time for realistic models \citep{Heggie_2014}.) Hence, one has also relied on statistical theories, especially kinetic theories. In the zeroth-order approximation $(N\to\infty)$, we may employ the `smooth' mean-field (MF) acceleration of star 1 due to $(N-1)$-body distribution function (DF) $F(2,\cdots,N,t)$
\begin{align}
\boldsymbol{A}_{1}(1,t)=\sum_{i=2}^{N}\int\text{d}2\cdots \text{d}N\, \boldsymbol{a}_{1i}\, F(2,\cdots,N)=\left(1-\frac{1}{N}\right)\int\text{d}2\, \boldsymbol{a}_{12}\,f(2,t)\approx\int\text{d}2\, \boldsymbol{a}_{12}\,f(2,t).\label{Eq.MF_A_two-body}
\end{align}
Throughout the present work, the following abbreviation is employed for the arguments of any dynamical functions and phase-space volume elements
\begin{subequations}
	\begin{align}
	&F(1,2,\cdots,N)=F(\boldsymbol{r}_{1},\boldsymbol{p}_{1},\boldsymbol{r}_{2},\boldsymbol{p}_{2},\cdots,\boldsymbol{r}_{N},\boldsymbol{p}_{N}),\\
	&\text{d}1\,\text{d}2\cdots\text{d}N=\text{d}\boldsymbol{r}_{1}\, \text{d}\boldsymbol{p}_{1}\,\text{d}\boldsymbol{r}_{2}\,\text{d}\boldsymbol{p}_{2}\cdots\text{d}\boldsymbol{r}_{N}\, \text{d}\boldsymbol{p}_{N},
	\end{align}\label{eq:abbrev}
\end{subequations}
where $(\boldsymbol{r}_{k},\boldsymbol{p}_{k})$ is the position-momentum phase-space point of star $k$ ($1\leq k\leq N$). In equation \eqref{Eq.MF_A_two-body}, we assume that the symmetry in permutation between the states of two stars and introduce the probable number DF $f(1,t)=NF(1,t)$ for the sake of self-consistency. 

The MF description is correct on time scales of dynamical time $t_\text{dyn}$($\sim0.1$ million years) on which the cluster may be in one of the quasi-stationary states due to the rapid change in MF potential \citep[e.g.,][]{Heggie_2003,Binney_2011}. On relaxation time scales $t_\text{rel}$($\sim 1$ giga year), stars may undergo a non-collective relaxation process due to the discreteness of the clusters, or the finiteness of total number $N$ (sometimes called finite-$N$ effect). Hence, the first-order approximation ($N\ll 1$) requires the actual acceleration to deviate from the MF acceleration of star 1
\begin{align}
\boldsymbol{a}_{1}-\boldsymbol{A}_{1}(1,t)=\sum^{N}_{i=2}\boldsymbol{a}_{1i}-\left(1-\frac{1}{N}\right)\int\text{d}2\, \boldsymbol{a}_{12}\,F(2,t).
\end{align}
The typical kinetic description relies on a two-body-DF description. One needs to simplify the acceleration of star 1 due to $(N-1)$-fields stars into the acceleration due to identical $(N-1)$-field stars (i.e., all the field stars are labeled as star 2) \citep[e.g.,][]{Kandrup_1981}
 \begin{align}
\boldsymbol{a}_{1}-\boldsymbol{A}_{1}(1,t)&\approx(N-1)\boldsymbol{a}_{12}-\left(1-\frac{1}{N}\right)\int\text{d}2\, \boldsymbol{a}_{12}\,f(2,t)=(N-1)\left(\boldsymbol{a}_{12}-\frac{1}{N}\int\text{d}2\, \boldsymbol{a}_{12}\,f(2,t)\right).
\end{align}
We define the \emph{`discreteness' stochastic acceleration} of star 1 as follows
\begin{align}
\tilde{\boldsymbol{a}}_{12}\equiv \boldsymbol{a}_{12}-\frac{1}{N}\int\text{d}2\, \boldsymbol{a}_{12}\,f(2,t).
\end{align}
The acceleration $\tilde{\boldsymbol{a}}_{12}$ corresponds to the ``discreteness fluctuation'' in the acceleration of star 1 due to the cluster's discreteness discussed in \citep{Kandrup_1988}. It was originally termed the ``statistical term'' in \citep{Gilbert_1968}.
The discreteness stochastic acceleration $\tilde{\boldsymbol{a}}_{12}$ satisfies the following relationship in the relaxation evolution of star clusters on secular-evolution time scales $t_\text{sec}\sim N\,t_\text{dyn}$ \citep[e.g.,][]{Kandrup_1988}
\begin{subequations}
\begin{align}
\left<\tilde{\boldsymbol{a}}_{12}\right>&=\int\text{d}2\,\tilde{\boldsymbol{a}}_{12}\,f(2,t)=0 \label{Eq.dis_fluc_null},\\
\dfrac{\left<\tilde{\boldsymbol{a}}_{12}\cdot\tilde{\boldsymbol{a}}_{12}\right>}{\left <\boldsymbol{A}\cdot\boldsymbol{A}\right>} &=\dfrac{\int\text{d}2\,\tilde{\boldsymbol{a}}_{12}\cdot\tilde{\boldsymbol{a}}_{12}\,f(2,t)}{\int\text{d}2\,\boldsymbol{A}_{1}\cdot\boldsymbol{A}_{1}\,f(2,t)}=\mathcal{O}\left(\frac{1}{N}\right)\neq 0, \label{Eq.dis_fluc_order}
\end{align}
\end{subequations}
where $<\cdot>$ means the averaging of a dynamical quantity with a proper stellar DF. However, the standard kinetic formalism relies on the method of characteristics for the correlation between particles \citep[e.g.,][]{Liboff_2003}, the following spatio-temporal correlation tensor of the stochastic acceleration is more convenient to use  (as given in equation \eqref{Eq.dsic_fluc})
\begin{align}
\ll\tilde{\boldsymbol{a}}_{12}(t)\,\tilde{\boldsymbol{a}}_{12}\left(t'\right)\gg=\iint\text{d}2(t)\,\text{d}2\left(t'\right)\,\tilde{\boldsymbol{a}}_{12}(t)\,\tilde{\boldsymbol{a}}_{12}(t')\,f(2(t),t)\,f\left(2\left(t'\right),t'\right),\hspace{1cm}(t>t')
\end{align}
where $2(t')$ are the phase space coordinates $(\boldsymbol{r}_{2},\boldsymbol{p}_{2})$ of star 2  at time $t'$ and $\tilde{\boldsymbol{a}}_{12}(t')$ is the discreteness stochastic acceleration of star 1 due to star 2 and the MF potential at time $t'$. In the rest of the present section, we review the discreteness stochastic acceleration of stars in the standard stellar dynamics (Section \ref{subsec:dsic_review}) and relatively recent kinetic theories (Section \ref{subsec:disc_kinetic}).

\subsection{Discreteness stochastic acceleration in the standard stellar dynamics and its importance}\label{subsec:dsic_review}
The order gap in equation \eqref{Eq.dis_fluc_order} infers that we consider only a `weakly-coupled' star cluster whose total potential energy is approximately $N$ times smaller than the total kinetic energy. The standard stellar-dynamics theory has examined the relaxation evolution of weakly-coupled star clusters using approximated forms of the discreteness stochastic acceleration of stars \citep{Spitzer_1988, Heggie_2003,Binney_2011}. Chandrasekhar's classical work \citep{Chandra_1942_a} discussed the stochastic nature of gravitational cumulative two-body relaxation in an infinite homogeneous system. Equation \eqref{Eq.dis_fluc_order} was approximated to the following form
\begin{align}
\ll\tilde{\boldsymbol{a}}_{12}(t)\,\tilde{\boldsymbol{a}}_{12}\left(t'\right)\gg_{\boldsymbol{A}\neq 0}\,\,\, \approx\,\,\, \ll\boldsymbol{a}_{12}(t)\,\boldsymbol{a}_{12}\left(t'\right)\gg_{\boldsymbol{A}=0}, \label{Eq.dis_fluc_approx_hom}
\end{align}
where the subscripts $\boldsymbol{A}\neq 0$ and $\boldsymbol{A}=0$ represent inhomogeneous and homogeneous distributions of field stars, respectively. \cite{Chandra_1943a} also employed the Holtsmark distribution of Newtonian-force strength to examine the stochastic many-body force exerted on a test star. Chandrasekhar showed that the force strength due to the nearest-neighbor stars around the test star dominated the vector sum of forces exerted on it (`dominant effect.'). Based on this consideration, the approximation in equation \eqref{Eq.dis_fluc_approx_hom}  results in a large gap between the relaxation-time and dynamical-time scales; 
\begin{align}
\frac{t_\text{rel}}{t_\text{dyn}}\approx\dfrac{\ll\tilde{a}_{12}(t)\,\tilde{a}_{12}\left(t'\right)\gg}{\ll A_{1}(t)\,A_{1}\left(t'\right)\gg}=\mathcal{O} \left( \frac{N}{\ln[N]}\right),\label{Eq.Coulomb_sec}
\end{align}
where $\tilde{a}_{12}$ and $A_{1}$ are the magnitudes of $\boldsymbol{\tilde{a}}_{12}$ and $\boldsymbol{A}_{1}$, respectively. The logarithmic factor ($\ln [N]$) in equation \eqref{Eq.Coulomb_sec} is called the `Coulomb logarithm.' However, as discussed  based on an analytical approach \citep{Cohen_1950} for classical plasmas and an $N$-body numerical simulation for finite inhomogeneous star clusters \citep{Aarseth_1998}, one must count the two-body relaxation due to distant stars/particles, in addition to the nearest stars/particles. Also, the relaxation time in equation \eqref{Eq.Coulomb_sec} is the outcome of a cumulative two-body relaxation that accumulates the effect of `independent' two-body relaxation on relaxation time scales. The independent relaxation is just an assumption. One must consider a many-body relaxation that distant stars simultaneously contribute to the relaxation process due to the Newtonian potential's long-range nature. 

Many-body relaxation is a `non-dominant effect' in the relaxation evolution of star clusters while it can help us understand the evolution, especially its early stage.
As seen from the expressions in equations \eqref{Eq.dis_fluc_order} and \eqref{Eq.Coulomb_sec},
the Coulomb logarithm is the order of unity and does not make a huge gap between the relaxation- and secular-evolution- times. Hence, one can modify the effect of two-body relaxation by including that of many-body relaxation, similarly to classical plasmas' relaxation process.\footnote{In plasma physics, the correct form of Coulomb logarithm has been discussed for the temperature relaxation of  classical ion-electron plasmas \citep[e.g.,][]{Gould_1967, Brown_2005}.} Of course, the correction to the two-body relaxation process is small. Hence, the many-body effect, or discreteness stochastic acceleration, is considered `non-dominant.' Also, discreteness stochastic acceleration can provide us with important traits of star clusters in the early stage of relaxation evolution, which the standard two-body-relaxation theory can not discuss. For example, discreteness stochastic acceleration could drive the relaxation evolution with an initial condition that stellar distances are close to the cluster-size scale  \citep{Kandrup_1981}. The stellar DF may become non-Maxwellian during its initial secular-evolution time if no evaporation and gravothermal instability occur \citep{Kandrup_1985}.  In fact, observed data showed that polytropic models (non-Maxwellian models) could well fit the structural profiles of low-concentration Galactic globular clusters with long relaxation times \citep{Ito_2020_3}. However, the direct relationship has not been clear between the non-Maxwellian DF and many-body relaxation. (We discuss the relationship in Section \ref{sec:import_discrete}.) 

\subsection{Discreteness stochastic acceleration in star-cluster kinetic theories}\label{subsec:disc_kinetic}
While many-body relaxation applies to all the star clusters (globular clusters and nuclear star clusters),  it has been rarely discussed even theoretically. To examine the many-body relaxation, one can resort to kinetic theories. \cite{Gilbert_1968} derived the currently most fundamental kinetic equation for $N$ stars in an isolated finite weakly-coupled star cluster, including the  gravitational polarization and the approximated discreteness stochastic acceleration. The former is a collective relaxation, and the latter is a non-collective one described by 
\begin{align}
\ll\tilde{\boldsymbol{a}}_{12}(t)\,\tilde{\boldsymbol{a}}_{12}\left(t'\right)\gg_{\boldsymbol{A}\neq 0} \,\,\,\approx\,\,\, \ll\boldsymbol{a}_{12}(t)\,\tilde{\boldsymbol{a}}_{12}\left(t'\right)\gg_{\boldsymbol{A}\neq 0}.
\end{align}
The fundamental feature of the discreteness stochastic acceleration has not been understood well and, one has not applied the concept to realistic models. This is  since the corresponding kinetic formulation is not complete yet. A correct kinetic formulation of the discreteness stochastic acceleration of stars without any approximation, $\left<\tilde{\boldsymbol{a}}_{12}(t)\,\tilde{\boldsymbol{a}}_{12}\left(t'\right)\right>_{\boldsymbol{A}\neq 0}$, was first rendered in \citep{Kandrup_1981}. Kandrup discussed the importance of the discreteness stochastic acceleration in the \emph{generalized Landau} (g-Landau) equation to describe the relaxation process for anisotropic stochastic systems \citep{Kandrup_1981_a} and the finite-$N$ effect on fluctuation in smooth MF potential force \citep{Kandrup_1988}. Those works are held only at the formal-expression level and can not explain the fundamental properties (\emph{H}-theorem, conservation laws, $\cdots$) of the g-Landau equation.

To understand a star-cluster kinetic equation's  fundamental properties in the relaxation time, one needs to employ an orbit-averaging of the kinetic equation and find the explicit expression in action-angle variables. The present work applies the orbit-averaging introduced in \citep{Polyachenko_1982} to the g-Landau kinetic equation. This approach is sensible when the effective pairwise interaction distance $r_\text{eff}$ is greater than the ``encounter radius'' \citep{Ogorodnikov_1965}. The radius is scaled as $\sim R/\sqrt{N}$ in a star cluster of dimension size $R$. Beyond the radius, the unperturbed motion of the test star can be determined dominantly by the MF Newtonian potential, and the periodicity of the motion can be secured.\footnote{The scaling of the encounter radius corresponds with that of the effective interaction distance in the Boltzmann-Grid limit of the standard gaseous system \citep[e.g.,][]{Cercignani_2008}. The encounter radius separates the random- and deterministic characteristics in the two-body collision. Cercignani's approach must be taken if we consider the effect of two-body relaxation properly.} We do not consider the orbit-averaging at $r_\text{eff}\lesssim R/\sqrt{N}$ to focus only on many-body relaxation.\footnote{If considering two-body relaxation at $r_\text{eff}\lesssim R/\sqrt{N}$, one must resort to the original orbit-averaging process \citep{Henon_1961}. H$\acute{\text{e}}$non's approach relies on the rectilinear trajectory of the test star's unperturbed motion and results in a FP kinetic equation whose collision terms include the Coulomb logarithm and assume homogeneous, local background stars.}  The Polyachenko's approach has been applied to the inhomogeneous Landau kinetic equation \citep{Polyachenko_1982,Luciani_1987,Chavanis_2008} and the inhomogeneous Balescu-Lanard kinetic equation \citep{Heyvaerts_2010,Chavanis_2012}. The orbit-averaging of the Gilbert's and g-Landau kinetic equations has not been discussed. The latter properly approximates the complicated mathematical structure of the Gilbert's equation, holding the demand of fundamental statistical mechanics (as we discuss in Section \ref{sec:Importance_gLandau}.) This feature can be shown by discussing correlation function's anti-normalization condition \citep{Liboff_1965,Liboff_1966}. The present paper aims to establish a kinetic formulation of the orbit-averaged g-Landau equation, showing the discreteness stochastic acceleration's significance in stellar statistical dynamics and its application to star clusters.

The present paper is organized as follows. Section \ref{sec:basic_kinetics} reviews a fundamental kinetic theory for the relaxation evolution of weakly-coupled star clusters and introduces the g-Landau kinetic equation. Section \ref{sec:Importance_gLandau}  shows that the g-Landau equation is only the kinetic equation that correctly approximates the \citep{Gilbert_1968}'s kinetic equation among the existing kinetic equation for the relaxation evolution, based on the anti-normalization condition. Section \ref{sec:explicit_orb_g_landau} shows the explicit form of the orbit-averaged g-Landau equation in action-angle variables. In Section \ref{sec:prop}, we reveal the basic properties of the g-Landau equation. Section \ref{sec:import_discrete} discusses the importance of the discreteness stochastic acceleration of stars in the early relaxation-evolution stage. It shows the  relationship of the many-body relaxation with non-Maxwellian stellar DF.  Section \ref{sec:conclusion} is Conclusion. 

\section{BBGKY hierarchy for the distribution function of stars in finite weakly-coupled star clusters}\label{sec:basic_kinetics}

Assuming that a star cluster is a finite, weakly-coupled self-gravitating system undergoing a relaxation evolution on secular-evolution time scales, one may mathematically model the time-evolution of the stellar DF by the Bogoliubov–Born–Green–Kirkwood–Yvon (BBGKY) hierarchy.  Section \ref{sec:BBGKY_rv} explains the BBGKY hierarchy in phase space $(\boldsymbol{r},\boldsymbol{p})$ and the anti-normalization condition for the stellar correlation function. Section \ref{sec:g_landau_rp} introduces the g-Landau kinetic equation in the phase space. 

\subsection{BBGKY hierarchy in phase space $(\textbf{r},\textbf{p})$ and anti-normalization condition}\label{sec:BBGKY_rv}

The Hamiltonian in terms of phase-space coordinates $(\boldsymbol{r}_{i},\boldsymbol{p}_{i})$ for $N$ stars of equal masses $m$ reads
\begin{equation}
\mathcal{H}=\sum_{i=1}^{N}\left(\frac{\boldsymbol{p}^{2}_{i}}{2m}+m\sum_{j>i}^{N}\phi(r_{ij})\right)=\sum_{i=1}^{N}\left(m\frac{\boldsymbol{v}^{2}_{i}}{2}-m\sum_{j>i}^{N}\frac{Gm}{r_{ij}}\right), \qquad(1\leq i, j \leq N \quad \text{with} \quad i\neq j )\label{Eq.H(r,p)}
\end{equation}
where $\boldsymbol{v}_{i}$ is the velocity of star $i$, and $r_{ij}=|\boldsymbol{r}_{i}-\boldsymbol{r}_{j}|$. Assuming that $N$ stars are statistically identical and indistinguishable, the $N$-body Liouville equation for the Hamiltonian $\mathcal{H}$ reduces to the BBGKY hierarchy \citep[e.g.,][]{Saslaw_1985,Landau_1987,Liboff_2003}  
\begin{align}
&\partial_{t}\, f_{s}(1,\cdots,s)+\sum_{i=1}^{s}\left[\boldsymbol{v}_{i}\cdot\frac{\partial}{\partial \boldsymbol{r}_{i}}+\sum_{\quad{j=1(\neq i)}}^{s}\boldsymbol{a}_{ij}\cdot\frac{\partial}{\partial \boldsymbol{v}_{i}}\right]f_{s}(1,\cdots,s)\nonumber\\
&\hspace{4cm}+\sum_{i=1}^{s}\left[\frac{\partial}{\partial \boldsymbol{v}_{i}}\cdot\int f_{s+1} (1,\cdots,s+1)\, \boldsymbol{a}_{i\,s+1}\, \text{d}(s+1)\right]=0,\label{Eq.BBGKY(r,p)}
\end{align}
where $\partial_{t}=\frac{\partial}{\partial t}$ and $1\leq s\leq N$. The function $f_{s}(1,\cdots,s)$ is the $s$-tuple DF, i.e., the probable number density of stars $1,\cdots,s$ that can be found at phase space points $(\boldsymbol{r}_{1},\boldsymbol{p}_{1}), \cdots,(\boldsymbol{r}_{s},\boldsymbol{p}_{s})$, respectively, at time $t$. To find a self-consistent kinetic equation for the single DF $f_{1}(1,t)$, one necessitates only the first two equations of the hierarchy for inhomogeneous, weakly-coupled finite clusters \citep[e.g.,][]{Gilbert_1971,Chavanis_2013a}
\begin{subequations}
	\begin{align}
	&(\partial_{t}+\boldsymbol{v}_{1}\cdot\nabla_{1})f_{1}(1,t)=-\boldsymbol{\partial}_{1}\cdot\int\text{d}2\, f_{2}(1,2,t)\, \boldsymbol{a}_{12},\label{Eq.1stBBGKY}\\
	&\left(\partial_{t}+\boldsymbol{v}_{1}\cdot\nabla_{1}+\boldsymbol{v}_{2}\cdot\nabla_{2}+\boldsymbol{a}_{12}\cdot\boldsymbol{\partial}_{12}\right)f_{2}(1,2,t)=-\int\text{d}3\,\left[\boldsymbol{a}_{13}\cdot\boldsymbol{\partial}_{1}+\boldsymbol{a}_{23}\cdot\boldsymbol{\partial}_{2}\right]f_{3}(1,2,3,t),\label{Eq.2nd_BBGKY}
	\end{align}
\end{subequations}
One may rewrite the single, double, and triple DFs by the correlation-function formalism based on the Mayer cluster expansion \citep[e.g.,][]{Mayer_1940,Green_1956}
\begin{subequations}
	\begin{align}
	f(1,t)\equiv& f_{1}(1,t),\label{Eq.single_DF}\\
	f(1,2,t)\equiv& f_{2}(1,2,t)= f(1,t)\,f(2,t)+\left[g(1,2,t)-\frac{f(1,t)\,f(2,t)}{N}\right],\label{Eq.double_DF}\\
	f(1,2,3,t)=&f(1,t)\,f(2,t)\,f(3,t)+\left(g(1,2,t)-\frac{f(1,t)\,f(2,t)}{N}\right)f(3,t)\nonumber\\ &+\left(g(2,3,t)-\frac{f(2,t)\,f(3,t)}{N}\right)f(1,t)+\left(g(3,1,t)-\frac{f(3,t)\,f(1,t)}{N}\right)f(2,t),\label{Eq.triple_DF}
	\end{align}
\end{subequations}
where $g(1,2)$, $g(1,3)$, and $g(2,3)$ are (binary) correlation functions. The ternary correlation function $T(1,2,3)$ is neglected so that encounters are rare events between binary and single stars. The DFs and correlation functions for stars may depend on the number $N$ as
\begin{subequations}
	\begin{align}
	&f(1,t),\, f(2,t),\, f(3,t)\sim N,\\
	&g(1,2,t),\, g(2,3,t),\, g(3,1,t)\sim N(N-1),\label{Eq.scaleg(1,2,t)}\\
	&G\sim1/N,\\
	&m,\,v,\,\boldsymbol{r},\,R,\,\boldsymbol{A}\sim 1,
	\end{align}\label{Eq.normalization}
\end{subequations}
where $R$ is the cluster size (i.e., the limiting/tidal radius or Jeans length of the cluster). In equation \eqref{Eq.normalization}, the normalization condition for the stellar DFs and correlation functions obeys  \citep{Liboff_1966}, and the scaling of physical quantities obeys \citep{Chavanis_2013a}. For weakly-coupled systems, the correlation-function formalism must satisfy the anti-normalization condition \citep{Liboff_1965}
\begin{align}
\int\text{d}1\, g(1,2,t)=0=\int \text{d}2\,g(1,2,t).\label{Eq.anti-normal}
\end{align}

\subsection{The g-Landau kinetic equation in phase space $(\textbf{r},\textbf{p})$}\label{sec:g_landau_rp}

If neglecting the effects of the gravitational polarization and close two-body encounter, one obtains the following two equations for DF $f(1,t)$ and correlation function $g(1,2,t)$ that compose the g-Landau equation \citep[See, e.g.,][]{Chavanis_2013a}
\begin{subequations}
	\begin{align}
	&\left(\partial_{t}+\boldsymbol{v}_{1}\cdot\nabla_{1}+\boldsymbol{A}_{1}\cdot\boldsymbol{\partial}_{1}\right)f(1,t)=-\boldsymbol{\partial}_{1}\cdot\int\text{d}2\, g(1,2,t)\, \tilde{\boldsymbol{a}}_{12},\label{Eq.1stBBGKY_gLandau}\\
	&\left(\partial_{t}+\boldsymbol{v}_{1}\cdot\nabla_{1}+\boldsymbol{v}_{2}\cdot\nabla_{2}+\boldsymbol{A}_{1}\cdot\boldsymbol{\partial}_{1}+\boldsymbol{A}_{2}\cdot\boldsymbol{\partial}_{2}\right)g(1,2,t)=-\left[\tilde{\boldsymbol{a}}_{12}\cdot\boldsymbol{\partial}_{1}+\tilde{\boldsymbol{a}}_{21}\cdot\boldsymbol{\partial}_{2}\right]f(1,t)\,f(2,t),\label{Eq.2ndBBGKY_gLandau}
	\end{align}
\end{subequations}
where the unperturbed trajectories of stars 1 and 2 in the Lagrangian description are
\begin{subequations}
	\begin{align}
	&\boldsymbol{r}_{i}(t)=\boldsymbol{r}_{i}(t-\tau)+\int_{t-\tau}^{t}\text{d}t'\,\boldsymbol{v}_{i}\left(t'\right),\hspace{15pt} (i=1\text{ or }2)\\
	&\boldsymbol{v}_{i}(t)=\boldsymbol{v}_{i}(t-\tau)+\int_{t-\tau}^{t}\text{d}t'\,\boldsymbol{A}_{i}\left(t'\right).
	\end{align}\label{Eq.characteristics_(r,p)}
\end{subequations}
The corresponding one-body Hamiltonian is
\begin{align}
h(\boldsymbol{r}_{i},\boldsymbol{p}_{i})=\frac{p^{2}_{i}}{2m}+\Phi(\boldsymbol{r}_{i},t)\equiv\frac{p^{2}_{i}}{2m}+\int\text{d}2\, \phi_{12}\,f(2,t). \qquad (i=1\text{ or }2)\label{Eq.one_body_Hamilton}
\end{align}

\section{Importance of generalized-Landau kinetic equation for the relaxation evolution of star clusters}\label{sec:Importance_gLandau}
Section \ref{sec:anti_norm} explains the anti-normalization condition for weakly-coupled star clusters in the kinetic formalism. Section \ref{sec:anti_def} examines the relationship of the anti-normalization condition with the definitions of the DF, the total number, and the total energy of stars. Section \ref{sec:anti_conserv} shows the importance of the anti-normalization condition to hold the conservation laws. We discuss these features based on the kinetic formalism in phase space $(\boldsymbol{r},\boldsymbol{p})$ \emph{before} operating an orbit-averaging of the BBGKY hierarchy. Lastly, we show that the g-Landau equation is a correct approximate of \citep{Gilbert_1968}'s equation from the point of view of fundamental statistical mechanics. We also show that the inhomogeneous Landau- and Balescu-Leanrd- equations are incorrect approximates of the Gilbert's equation. 

\subsection{Anti-normalization condition of the g-Landau equation at the equation level}\label{sec:anti_norm}

We explain the antinormalization condition for \citep{Gilbert_1968}'s equation, inhomogeneous Landau equation \citep{Polyachenko_1982}, inhomogeneous Balescu-Lenard equation \citep{Heyvaerts_2010}, and the g-Landau equation \citep{Kandrup_1981}.  The first two equations in the BBGKY hierarchy corresponding to the star-cluster kinetic equations read
\begin{subequations}
	\begin{align}
	&\left(\partial_{t}+\boldsymbol{v}_{1}\cdot\nabla_{1}+\boldsymbol{A}_{1}\cdot\boldsymbol{\partial}_{1}\right)f(1,t)=-\boldsymbol{\partial}_{1}\cdot\int\text{d}2\, g(1,2,t)\, \boldsymbol{a}_{12},\label{Eq.1stBBGKY_Gilbert}\\
	&\left(\partial_{t}+\boldsymbol{v}_{1}\cdot\nabla_{1}+\boldsymbol{v}_{2}\cdot\nabla_{2}+\boldsymbol{A}_{1}\cdot\boldsymbol{\partial}_{1}+\boldsymbol{A}_{2}\cdot\boldsymbol{\partial}_{2}\right)g(1,2,t)=C_{c}, \label{Eq.2ndBBGKY_Gilbert}
	\end{align}
\end{subequations}
where
\begin{align}
C_{c}=\begin{cases}-\left[\boldsymbol{a}_{12}\cdot \boldsymbol{\partial}_{1}+\boldsymbol{a}_{21}\cdot\boldsymbol{\partial}_{2}\right]f(1,t)f(2,t),\hspace{6.6cm} (\text{Landau})\\
-\left[\boldsymbol{\tilde{a}}_{12}\cdot \boldsymbol{\partial}_{1}+\boldsymbol{\tilde{a}}_{21}\cdot\boldsymbol{\partial}_{2}\right]f(1,t)f(2,t),\hspace{6.3cm} (\text{g-Landau})\\
-\left[\boldsymbol{a}_{12}\cdot \boldsymbol{\partial}_{1}+\boldsymbol{a}_{21}\cdot\boldsymbol{\partial}_{2}\right]f(1,t)f(2,t)\,\\
\hspace{1.3cm}-\int\text{d}3\,g(2,3,t)\,\boldsymbol{a}_{13}\cdot\boldsymbol{\partial}_{1}f(1,t)-\int\text{d}3\,g(1,3,t)\,\boldsymbol{a}_{23}\cdot\boldsymbol{\partial}_{2}f(2,t),\hspace{0.3cm} (\text{Balescu-Lenard})\\
-\left[\boldsymbol{\tilde{a}}_{12}\cdot \boldsymbol{\partial}_{1}+\boldsymbol{\tilde{a}}_{21}\cdot\boldsymbol{\partial}_{2}\right]f(1,t)f(2,t)\,\\
\hspace{1.7cm}-\int\text{d}3\,g(2,3,t)\,\boldsymbol{a}_{13}\cdot\boldsymbol{\partial}_{1}f(1,t)-\int\text{d}3\,g(1,3,t)\,\boldsymbol{a}_{23}\cdot\boldsymbol{\partial}_{2}f(2,t).\hspace{1.2cm} (\text{Gilbert})\label{Eq.Cc}
\end{cases}
\end{align}
Assume that the anti-normalization condition is satisfied in equation \eqref{Eq.2ndBBGKY_Gilbert}. By taking the integral $\int \cdot\,\,\text{d}1$ (or $\int \cdot\,\,\text{d}2$) over each side of the equation and employing the boundary conditions for the DF and correlation function
\begin{subequations}
	\begin{align}
	&\int\text{d}\boldsymbol{v}_{1}\,\boldsymbol{\partial}_{1}\, f(1,t)=0, \hspace{2cm}\int\text{d}\boldsymbol{v}_{1}\,\boldsymbol{\partial}_{1}\, g(1,2,t)=0,\\ 
	&\int\text{d}\boldsymbol{r}_{1}\, \nabla_{1}\, f(1,t)=0,\hspace{2cm}\int\text{d}\boldsymbol{r}_{1}\, \nabla_{1}\, g(1,2,t)=0,
	\end{align}
\end{subequations}
one can easily confirm that all the terms on the left-hand side of equation \eqref{Eq.2ndBBGKY_Gilbert} vanish while obtaining, on the right-hand side,
\begin{align}
    \int\text{d}2\, C_{c}=\begin{cases}
    -\boldsymbol{A}_{1}\cdot \boldsymbol{\partial}_{1}f(1,t),\hspace{1.0cm}(\text{Landau and Balescu-Lenard})\\
    0. \hspace{3.2cm} (\text{Gilbert's and g-Landau})
    \end{cases}
\end{align}
Hence, the inhomogeneous Landau- and Balescu-Lenard- equations are inconsistent. The reason why the Gilbert's and g-Landau equations satisfy the anti-normalization condition at the equation level is due to the existence of the term $\left[\boldsymbol{A}_{1}\cdot \boldsymbol{\partial}_{1}+\boldsymbol{A}_{2}\cdot\boldsymbol{\partial}_{2}\right]f(1,t)f(2,t)$ in $C_{c}$. The term cancels out the term $-\boldsymbol{A}_{1}\cdot \boldsymbol{\partial}_{1}f(1,t)$ in $\int \text{d}2\,C_{c}$.

\subsection{Anti-normalization condition and the definitions for the total energy and the total number of stars}\label{sec:anti_def}

Excluding the effect of strong-close encounters, the kinetic equation derived in \citep{Gilbert_1968} is the most accurate description at the two-body-DF truncation level. However, its physical feature and application have not been fully understood due to the complex mathematical structure. One needs a fundamental criterion to approximate the equation. The present work relies on the anti-normalization condition. Since the derivation of the \citep{Gilbert_1968}'s equation is based on the first two equations of the BBGKY hierarchy with three body DF neglecting the ternary correlation function, the definition of the total number of stars in a weakly-coupled star cluster is
\begin{subequations}
	\begin{align}
	N(t)&=\int\cdots\int\text{d}1\cdots\, \text{d}N\,\sum_{i=1}^{N}F_{N}(1\cdots N,t),\\
	&=N\int\text{d}1\, \text{d}2\, \text{d}3\left[f(1,t)f(2,t)f(3,t)+g(1,2,t)f(3,t)+g(2,3,t)f(1,t)+g(3,1,t)f(2,t)\right],\label{Def.N3}
	\end{align}
\end{subequations}
and the total energy is
\begin{subequations}
\begin{align}
E(t)&=\int\cdots\int\text{d}1\cdots\text{d}N\,\sum_{i=1}^{N}\left(\frac{\boldsymbol{p}^{2}_{i}}{2m}+m\sum_{j>i}^{N}\phi(r_{ij})\right)F_{N}(1\cdots N,t),\\
&=\int\text{d}1\, \frac{\boldsymbol{p}_{1}^{2}}{2m}f_{1}(1,t)\, +U_{\text{m.f.}}(t)+U_{\text{cor}}(t),\label{Eq.E}\\
U_{\text{m.f.}}(t)&=\frac{m}{2}\int\text{d}1\,\Phi(\boldsymbol{r}_{1},t)\,f(1,t),\label{Eq.U_id}\\
U_{\text{cor}}(t)&=\frac{m}{2}\left(\iint\text{d}{1}\,\text{d}{2}\,\phi(r_{12})\,g(1,2,t)+\iiint\text{d}{1}\,\text{d}{2}\,\text{d}3\,\phi(r_{12})\left[f(1,t)\,g(2,3,t)+f(2,t)\,g(3,1,t)\right]\right),\label{Eq.U_cor}
\end{align}\label{Def.E3}
\end{subequations}
If the anti-normalization condition holds, we retrieve the standard expressions\footnote{The integral $\iint\text{d}1\, \text{d}2\,  g(1,2,t)\, \phi(r_{12})$ in the total energy is typically neglected since it is $N$ times less significant than the rest of the terms. However, the present work leaves the term in the total energy since the Gilbert's and g-Landau- equations are derived by expanding the BBGKY hierarchy by the discreteness parameter, $1/N$.} of the total number and the total energy
\begin{align}
N(t)=\int\text{d}1\, f(1,t),\qquad\qquad E(t)=\int\text{d}1\, \frac{\boldsymbol{p}_{1}^{2}}{2m}\, f(1,t)+U_\text{m.f.}+\iint\text{d}1\, \text{d}2\,  g(1,2,t)\, \phi(r_{12}).\label{Def.N2_E2}
\end{align}
The anti-normalization condition is important to correctly define the DF, the total energy, and the total number of stars at the order of unity in equation \eqref{Def.N2_E2}. If a kinetic formulation can not satisfy the condition, one must validate the the definitions of total number and total energy, equations \eqref{Def.N3} and \eqref{Def.E3}. (In equation \eqref{Def.E3}, even the polarization effect contributes to the total energy!) 

We can easily confirm in equation \eqref{Eq.U_cor} that employing the anti-normalization condition is equivalent to neglecting the gravitational polarization effect $\left(\int\text{d}3\,\left[\phi(r_{13})\,g(2,3,t)+\phi(r_{23})\,g(1,3,t)\right]\right)$. As long as a kinetic equation satisfies the anti-normalization condition, one may neglect the effect of the polarization as an approximation, which still holds the standard definitions of the total number and the total energy of stars.

\subsection{Anti-normalization condition and the conservation laws of  total energy and total number for the g-Landau equation}\label{sec:anti_conserv}

For completeness, we first explain how we can prove the conservation of the total energy and the total number of stars for \citep{Gilbert_1968}'s equation. We can prove the conservation of total number, $\text{d}N(t)/\text{d}t=0$, simply by taking the integral $\int\cdot\,\,\text{d}1$ over each side of equation \eqref{Eq.1stBBGKY_Gilbert} and using the definition of total number, equation \eqref{Def.N2_E2}. To prove the conservation of total energy, $\text{d}E(t)/\text{d}t=0$, we must find the time-evolution equation for $f(1,2,t)$, employing equations \eqref{Eq.1stBBGKY_Gilbert} and  \eqref{Eq.2ndBBGKY_Gilbert} as follows
\begin{align}
&\left(\partial_{t}+\boldsymbol{v}_{1}\cdot\nabla_{1}+\boldsymbol{v}_{2}\cdot\nabla_{2}+\boldsymbol{A}_{1}\cdot\boldsymbol{\partial}_{1}+\boldsymbol{A}_{2}\cdot \boldsymbol{\partial}_{2}\right)f(1,2,t)\nonumber\\
&\hspace{1.0cm}=-\int\text{d}3\, g(2,3,t)\, \boldsymbol{a}_{13}\cdot\boldsymbol{\partial}_{1}f(1,t) -\int\text{d}3\, g(1,3,t)\, \boldsymbol{a}_{23}\cdot\boldsymbol{\partial}_{2}f(2,t)\nonumber\\
&\hspace{1.4cm}-\left[\tilde{\boldsymbol{a}}_{12}\cdot\boldsymbol{\partial}_{1}+\tilde{\boldsymbol{a}}_{21}\cdot\boldsymbol{\partial}_{2}\right]f(1,t)f(2,t)-\boldsymbol{\partial}_{1}\cdot\int\text{d}3\, g(1,3,t)\, \boldsymbol{a}_{13}-\boldsymbol{\partial}_{2}\cdot\int\text{d}3\, g(2,3,t)\, \boldsymbol{a}_{23}.\label{Eq.2ndBBGKY_Gilbert_f(1,2,t)}
\end{align}
By multiplying equations \eqref{Eq.1stBBGKY_Gilbert} and \eqref{Eq.2ndBBGKY_Gilbert_f(1,2,t)} by the factors $\frac{\boldsymbol{p}_{1}^{2}}{2m}$ and $\phi_{12}$, respectively, then adding the two equations up, one can prove that the total energy can conserve. In the processes above to prove the conservation laws, the terms $\left(\int\text{d}3\,g(2,3,t)\, \boldsymbol{a}_{13}\cdot\boldsymbol{\partial}_{1}f(1)+\int\text{d}3\,g(1,3,t)\, \boldsymbol{a}_{23}\cdot\boldsymbol{\partial}_{2}f(2)\right)$ associated with the gravitational polarization do not come into play. Hence, the same result can be true for the g-Landau equation as well. On the other hand, one must prove the conservation of  total energy and total number for the inhomogeneous Balescu-Lenard equation and inhomogeneous Landau equation, using equations \eqref{Def.N3} and \eqref{Def.E3}, if possible. 

 Table \ref{table:Equations} shows the summary of the anti-normalization condition and the conservation laws for the Gilbert's, g-Landau-, inhomogeneous Balescu-Lenard-, and inhomogeneous-Landau equations. As one sees a kinetic equation listed lower on the table, total energy and total number deviate more largely from their correct definitions. Only the g-Landau equation can satisfy all the conservation laws and the anti-normalization condition at both the kinetic-equation level and the explicit-form level (as we further prove the latter in Section \ref{subsec.conserv}.) As a reference, we also list the most often-employed heuristic star-cluster kinetic equation, i.e., homogeneous, local Fokker-Planck (FP) kinetic equation. One can not consistently define the total energy for the FP equation since the equation does not count the MF potential self-consistently. On the other hand, one can define the total stellar number  correctly  since the FP equation satisfies the anti-normalization even at the explicit-form level (See Appendix \ref{Appendix:anti}). The  present section concludes that only the g-Landau equation is a correct approximate of the \citep{Gilbert_1968}'s equation since it satisfies the anti-normalization condition. Accordingly, it can correctly define the DF, the total energy, and the total number of stars, satisfying the conservation laws.

\begin{table}\centering
	\scalebox{0.8}{
	\begin{tabular}{l l l ll l}
		\hline
		Kinetic equations & Total number & Total energy & Anti-normalization&Refs\\
                                      &                   &                  &in $(\boldsymbol{r},\boldsymbol{p})$& \\
		\hline
		Gilbert's equation & conserved & conserved&  hold&\citep{Gilbert_1968}, \citep{Gilbert_1971}\\
               (d.s.a., g.p.)          &                 &               &        &           \\
             \hline
             g-Landau equation & conserved & explicitly&  explicitly hold&\citep{Kandrup_1981}, \citep{Kandrup_1988},\\
               (d.s.a., - )                     & &  conserved             &        & \citep{Chavanis_2013b}, (\textbf{the present work})          \\
             \hline
             inhomogeneous Landau & exceed by  & exceed by  & fail&\citep{Polyachenko_1982}, \\
               ( - , - )                      &  $\Delta N_{g}$               &   $-\Phi(1)/N$            &        &  \citep{Luciani_1987}          \\
             \hline
             inhomogeneous Balescu-Lenard & exceed by  & exceed by  &  fail& \citep{Heyvaerts_2010}, \citep{Chavanis_2012}\\
               ( - , g.p.)          &  $\Delta N_{g}$               &   $-\Phi(1)/N+U_{p}$            &        &           \\
             \hline
             local/homogeneous FP & conserved&  undefined &  explicitly hold& \citep{Rosenbluth_1957}, \citep{Henon_1961},\\
                ( - , - )                      &                 &               &        & \citep{Cohn_1979}         \\
		\hline
	\end{tabular}
}
\caption{Anti-normalization condition and the conservation of the total energy and the total number of stars for star-cluster kinetic equations. The letters (d.s.a., g.p.) are added if a kinetic equation includes the effect(s) of the discreteness stochastic acceleration (d.s.a.) and/or gravitational polarization (g.p.). The excesses of total energy and total number are defined as $\Delta N_{g}$ $=$ $N\iiint\text{d}1\,\text{d}2\,\text{d}3\,$ $[g(1,2,t)\,f(3,t)$ $+$ $g(2,3,t)\,f(1,t)$ $+$ $g(3,1,t)\,f(2,t)]$ and $U_{p}$ $=$ $\frac{m}{2}$ $\iiint\text{d}{1}\,\text{d}{2}\,\text{d}3$ $\,\phi(r_{12})$ $[f(1,t)\,g(2,3,t)$ $+$ $f(2,t)\,g(3,1,t)]$. In the column for the anti-normalization condition, the term "hold/fail" means that the equation is/is-not consistent after the integral $\int \cdot\,\,\text{d}1$ or $\int \cdot\,\,\text{d}2$ is taken over each side of the second equation of the BBGKY hierarchy.
The term "explicitly conserved/hold" is used when the explicit expression of a physical quantity satisfies the corresponding conservation law or when that of the correlation function satisfies the anti-normalization condition. The references are related to the derivation of the kinetic equations and the orbit-averaged equation, while only the orbit-averaging of the Gilbert's equation has not been done yet.}
\label{table:Equations}
\end{table}

\section{The explicit expression of the orbit-averaged generalized-Landau kinetic equation in action-angle space}\label{sec:explicit_orb_g_landau}

In Section \ref{sec:BBGHY_action_angle}, we show the first two equations of an orbit-averaged BBGKY hierarchy in action-angle variables. Then, Section \ref{sec:fLandau_action_angle} provides the explicit expressions of the collision terms and the orbit-averaged g-Landau equation.

\subsection{Orbit-averaging of the BBGKY hierarchy in action-angle variables}\label{sec:BBGHY_action_angle}

In the secular evolution of a star cluster, stars may be in a quasi-stationary state on dynamical time scales. This state is mathematically modeled as the solution of the time-independent collisionless-Boltzmann (Vlasov) equation 
\begin{align}
	&\left(\boldsymbol{v}_{1}\cdot\nabla_{1}+\boldsymbol{A}_{1}\cdot\boldsymbol{\partial}_{1}\right)f(1,t)=[f(1,t),h(\boldsymbol{r}_{1},\boldsymbol{v}_{1})]_{(r_{1},v_{1})}=0, \qquad\qquad (t\sim t_\text{dyn}, \quad \text{or}, \quad N\to \infty) \label{Eq.t_ind_Vlasov}
	\end{align}
	where the Poisson bracket for dynamical functions $A(\boldsymbol{r},\boldsymbol{v})$ and $B(\boldsymbol{r},\boldsymbol{v})$ is defined as
\begin{align}
[A,B]_{(r,v)}=\frac{\partial A}{\partial \boldsymbol{r}}\cdot\frac{\partial B}{\partial \boldsymbol{v}}-\frac{\partial A}{\partial \boldsymbol{v}}\cdot\frac{\partial B}{\partial \boldsymbol{r}}.
\end{align}
If the orbits of stars are arguably regular, and the MF potential is unchanged with time, the strong Jeans theorem \citep[e.g.,][]{Binney_2011} states that the solution to equation \eqref{Eq.t_ind_Vlasov} is any function of action vector, $\boldsymbol{I}_{1}\equiv(I_{1\alpha},I_{1\beta},I_{1\gamma})$,
\begin{align}
f(1,t)=f(\boldsymbol{I}_{1}). \qquad\qquad (\text{if}\,\,\,\,t\sim t_\text{dyn}, \,\,\,\, \text{or}\,\,\,\, N\to \infty) \label{Eq.f(I)}
\end{align}
Assuming that the Hamiltonian is integrable in equation \eqref{Eq.one_body_Hamilton}, one may convert the Hamiltonian equations under the canonical transformation from phase-space coordinates to action-angle ones
\begin{align}
&\frac{\text{d}\boldsymbol{\omega}}{\text{d}t}=\frac{\partial h}{\partial \boldsymbol{I}}\equiv\Omega(\boldsymbol{I})\qquad\text{ and }\qquad \frac{\text{d}\boldsymbol{I}}{\text{d}t}=\frac{\partial h}{\partial \boldsymbol{\omega}}=0.\label{Eq.Hamilton_Jacobi_eqns}
\end{align}
Accordingly,
\begin{align}
&\boldsymbol{\omega}=\boldsymbol{\Omega}\,t+\boldsymbol{\omega}_\text{o}\qquad\text{ and }\qquad h=h(\boldsymbol{I}),\label{Eq.Hamilton_eqns_Iw}
\end{align}
where $\boldsymbol{\omega}_\text{o}$ is a constant angle vector. The Hamiltonian equation \eqref{Eq.Hamilton_eqns_Iw} satisfies the conservation of the one-body Hamiltonian
\begin{align}
\frac{\text{d}h}{\text{d}t}=\boldsymbol{\omega}\cdot\frac{\text{d}\boldsymbol{I}}{\text{d}t}=0.\label{Eq.h_cons}
\end{align}

The discussion above must be modified due to the cluster's finite-$N$ (discreteness) effect on secular-evolution time scales. The action vector may depend on time $t$, and the DF is also a function of the action angles $\boldsymbol{\omega}$
\begin{align}
f(1,t)\approx f(\boldsymbol{I}_{1}(t),t)+\frac{f(\boldsymbol{I}_{1}(t),\boldsymbol{\omega}_{1},t,t/N)}{N}+\mathcal{O}\left(\frac{1}{N^{2}}\right), \qquad (\text{if}\,\,\,\,t\sim t_\text{rel},\,\,\,\, \text{or} \,\,\,\, N \text{ is finite.})
\end{align}
However, one conventionally assumes that the one-body Hamiltonian $h(\boldsymbol{I}_{1})$ is still integrable and that the action vector is time-independent even on secular-evolution time scales against the correct one \citep[e.g.,][]{Henon_1961}.\footnote{Existing works for the inhomogeneous star-cluster kinetic equations (in Table \ref{table:Equations}) conventionally employ a stronger assumption than  \citep{Henon_1961}. \cite{Henon_1961} considered in the local, homogeneous FP equation that the phase space volume (or the radial action integral) changes with time on relaxation time scales. Hence, one can employ a self-similar analysis on the FP equation \citep{Heggie_1988,Takahashi_1993,Ito_2021} while not for the inhomogeneous kinetic equations, including the g-Landau equation.}  Hence, one typically considers that equations \eqref{Eq.f(I)} - \eqref{Eq.h_cons} can hold in the secular evolution \citep[e.g.,][]{Polyachenko_1982,Heyvaerts_2010}. This treatment mathematically enables us to avoid the nonlinearity in the trajectory of stars, equation \eqref{Eq.characteristics_(r,p)}. On the one hand, employing the action-angle approach costs a self-consistency in energy conservation. The total energy, equation \eqref{Def.N2_E2}, can not be conserved because of equation \eqref{Eq.h_cons} (as we show in Section \ref{subsec.conserv}.). Also, the `target' of the action-angle approach is limited to the motion of stars only on scales larger than the encounter radius $(\sim R/\sqrt{N})$. Beyond the radius, the MF potential dominates the motion of stars and secures the periodicity of stars' motion. Accordingly, the approach can not correctly count the effect of encounters close to the Landau length $(\sim R/N)$ \citep[See, e.g.,][for the Landau length]{Landau_1936,Spitzer_1988,Chavanis_2013a}. Hence, it appears proper to employ an orbit-averaged kinetic equation with the many-body relaxation, excluding the two-body relaxation, for applications. (We extend this discussion in Section \ref{sec:import_discrete}.) Also, one can not obtain a simple expression of collision terms (and relaxation time) from the action-angle approach. Rather, one obtains a summation of infinite series with a complex expression of integrals including delta functions (as we obtain in Section \ref{subsec:glanda_WI}). 

Assume that the one-body Hamiltonian $h(\boldsymbol{I})$ is autonomous in the secular evolution of clusters. Then, one may rewrite the first two equations of the BBGKY hierarchy, equation \eqref{Eq.2ndBBGKY_Gilbert}, in action-angle variables. Employing the Poisson bracket's invariance under the canonical transformation, the equations with the g-Landau collision terms reduce to
\begin{subequations}
	\begin{align}
	&\partial_{t}f(1,t)+[f(1,t),h(1)]_{(\omega_{1},I_{1})}=\int\text{d}{2}\,\left[\phi_{12},g(1,2,t)\right]_{(\omega_{1},I_{1})},\label{Eq.1stBBGKY_gLandau_brac}\\
	&\partial_{t}g(1,2,t)+[g(1,2,t),h(1)]_{(\omega_{1},I_{1})}+[g(1,2,t),h(1)]_{(\omega_{2},I_{2})}\nonumber\\
	&\hspace{2.0cm}=\sum_{i=1}^{2}\left\{[\phi_{12},f(1,t)\,f(2,t)]_{(\omega_{i},I_{i})}-\frac{1}{N}\int\text{d}3\,\left[\phi_{i3},f(1,t)\,f(2,t)\,f(3,t)\right]_{(\omega_{i},I_{i})} \right\},\label{Eq.2ndBBGKY_gLandau_brac}
	\end{align}\label{Eq.BBGKY_gLandau_brac}
\end{subequations}
where we use the abbreviation $f(1,t)=f(\boldsymbol{I}_{1},\boldsymbol{\omega}_{1},t)$,  $g(1,2,t)=g(\boldsymbol{I}_{1},\boldsymbol{\omega}_{1},\boldsymbol{I}_{2},\boldsymbol{\omega}_{2},t)$,  $\text{d}2=\text{d}\boldsymbol{I}_2\,\text{d}\boldsymbol{\omega}_2$, and $\text{d}3=\text{d}\boldsymbol{I}_3\,\text{d}\boldsymbol{\omega}_3$. (We believe these abbreviations do not cause  confusion, compared with the corresponding phase-space description in equation \eqref{eq:abbrev}.)  To derive equation \eqref{Eq.BBGKY_gLandau_brac}, the invariance of phase space volumes elements ($\text{d}\boldsymbol{r}_{2}\,\text{d}\boldsymbol{v}_{2}=\text{d}\boldsymbol{I}_{2}\,\text{d}\boldsymbol{\omega}_{2}$ and $\text{d}\boldsymbol{r}_{3}\,\text{d}\boldsymbol{v}_{3}=\text{d}\boldsymbol{I}_{3}\,\text{d}\boldsymbol{\omega}_{3}$) is also employed. If one assumes that the DF $f(1,t)$ depends only on the action vector (the strong Jeans theorem) and takes the orbit-averaging,  $\int\cdot\,\,\text{d}\boldsymbol{\omega}_{1}/(2\pi)^{3}$, only over equation \eqref{Eq.1stBBGKY_gLandau_brac}, then the corresponding two equations read
\begin{subequations}
	\begin{align}
	&\partial_{t}f(\boldsymbol{I}_{1},t)=C\left[f(\boldsymbol{I}_{1},t)\right],\label{Eq.1stBBGKY_gLandau_orb}\\
	&\partial_{t}g(1,2,t)+\boldsymbol{\Omega}_{1}\cdot\frac{\partial g(1,2,t)}{\partial \boldsymbol{\omega}_{1}}+\boldsymbol{\Omega}_{2}\cdot\frac{\partial g(1,2,t)}{\partial \boldsymbol{\omega}_{2}}\nonumber\\
&\hspace{4cm}=\sum_{i=1}^{2}\frac{\partial }{\partial \boldsymbol{\omega}_{i}}\left(\phi_{12}-\frac{1}{N}\int\text{d}3\,\phi_{i3}\,f(\boldsymbol{I}_{3},t)\right)\cdot\frac{\partial f(\boldsymbol{I}_{1},t)\,f(\boldsymbol{I}_{2},t)}{\partial \boldsymbol{I}_{i}} ,\label{Eq.2ndBBGKY_gLandau_orb}
	\end{align}
\end{subequations}
where the collision term $C\left[f(\boldsymbol{I},t)\right]$ is
\begin{align}
C\left[f(\boldsymbol{I}_{1},t)\right]\equiv\iint\frac{\text{d}2\,\text{d}\boldsymbol{\omega}_{1}}{8\pi^{3}}\left(\frac{\partial \phi_{12}}{\partial \boldsymbol{\omega}_{1}}\cdot\frac{\partial g(1,2,t)}{\partial \boldsymbol{I}_{1}}-\frac{\partial \phi_{12}}{\partial \boldsymbol{I}_{1}}\cdot\frac{\partial g(1,2,t)}{\partial \boldsymbol{\omega}_{1}}\right).\label{Eq.collision_gLandau_orb}
\end{align}

\subsection{The explicit expression of the orbit-averaged generalized Landau equation in action-angle variables}\label{sec:fLandau_action_angle}

Sections \ref{subsec:corr} and \ref{subsec:glanda_WI} find the explicit expression of both the correlation function and the collision integrals for the orbit-averaged g-Landau equation.

\subsubsection{Correlation function in action-angle variables}\label{subsec:corr}

To find a self-consistent kinetic equation, we solve the second equation of the orbit-averaged BBGKY hierarchy, equation \eqref{Eq.2ndBBGKY_gLandau_orb}, using the method of characteristics along the unperturbed trajectory described by the Hamiltonian equation \eqref{Eq.Hamilton_eqns_Iw}
\begin{subequations}
	\begin{align}
	&g(1,2,t)=g_{a}(1,2,t)+g_{A}(1,2,t),\\
	&g_{a}(1,2,t)\equiv\sum_{i=1}^{2}\int^{t}_{t-\tau}\text{d}t'\,\left.\frac{\partial \phi_{12}}{\partial \boldsymbol{\omega}_{i}}\right|_{t=t'}\cdot\frac{\partial }{\partial \boldsymbol{I}_{i}(t')}\left[f\left(\boldsymbol{I}_{1}(t'),t'\right)\, f\left(\boldsymbol{I}_{2}(t'),t'\right)\right],\\
	&g_{A}(1,2,t)\equiv -\frac{1}{N}\sum_{i=1}^{2}\int^{t}_{t-\tau}\text{d}t'\int\text{d}3\,\left.\frac{\partial \phi_{i3}}{\partial \boldsymbol{\omega}_{i}}\right|_{t=t'}\cdot\frac{\partial}{\partial \boldsymbol{I}_{i}(t')}\left[f(\boldsymbol{I}_{1}(t'),t')\, f(\boldsymbol{I}_{2}(t'),t')\, f(\boldsymbol{I}_{3},t)\right],\label{Eq.g_A}
	\end{align}
\end{subequations}
where only the arguments associated with star 1 and star 2 are 'moved' with time, obeying each trajectory. Accordingly, we separate the collision term $C\left[f(\boldsymbol{I}_{1},t)\right]$ of equation \eqref{Eq.collision_gLandau_orb} as follows
\begin{align}
    C\left[f(\boldsymbol{I}_{1},t)\right]=C_{a}\left[f(\boldsymbol{I}_{1},t)\right]+C_{A}\left[f(\boldsymbol{I}_{1},t)\right],
\end{align}
where $C_{a}$ and $C_{A}$ are the collision terms attributed to $g_{a}(1,2,t)$ and $g_{A}(1,2,t)$, respectively. The explicit expression of the correlation function $g_{a}(1,2,t)$ has already been derived in \citep{Polyachenko_1982} as follows
\begin{align}
g_{a}(1,2,t)=&i\pi \sum_{\boldsymbol{n}_{1}=-\infty}^{\infty}\sum_{\boldsymbol{n}_{2}=-\infty}^{\infty} A_{\boldsymbol{n}_{1}\boldsymbol{n}_{2}}(\boldsymbol{I}_{1},\boldsymbol{I}_{2})\, \delta(\boldsymbol{n}_{1}\cdot\boldsymbol{\Omega}_{1}-\boldsymbol{n}_{2}\cdot\boldsymbol{\Omega}_{2})\, e^{i\left(\boldsymbol{n}_{1}\cdot\boldsymbol{\omega}_{1}-\boldsymbol{n}_{2}\cdot\boldsymbol{\omega}_{2}\right)}\nonumber\\
&\hspace{3cm}\times\left[\boldsymbol{n}_{1}\cdot\frac{\partial}{\partial \boldsymbol{I}_{1}}-\boldsymbol{n}_{2}\cdot\frac{\partial}{\partial \boldsymbol{I}_{2}}\right]f(\boldsymbol{I}_{1},t)f(\boldsymbol{I}_{2},t).\label{Eq.g_a(1,2,t)}
\end{align}
Hence, our focus is the correlation function $g_{A}(1,2,t)$ only. One can Fourier-transform the potentials $\phi_{13}$ and $\phi_{23}$ with respect to the action-angle variables
\begin{subequations}
	\begin{align}
	&\phi_{k3}=\sum_{n_{k}=-\infty}^{\infty}\sum_{n_{3}=-\infty}^{\infty}A_{\boldsymbol{n}_{k}\boldsymbol{n}_{3}}(\boldsymbol{I}_{k},\boldsymbol{I}_{3})\,e^{i(\boldsymbol{n}_{k}\cdot\boldsymbol{\omega}_{k}-\boldsymbol{n}_{3}\cdot\boldsymbol{\omega}_{3})},\qquad (k=1\text{ or } 2)\label{Eq.inv_Fourier_phi}\\
	&A_{\boldsymbol{n}_{k}\boldsymbol{n}_{3}}(\boldsymbol{I}_{k},\boldsymbol{I}_{3})=\frac{1}{(8\pi^{3})^{2}}\iint\text{d}\boldsymbol{\omega}_{k}\, \text{d}\boldsymbol{\omega}_{3}\, \phi_{k3} \,e^{-i(\boldsymbol{n}_{k}\cdot\boldsymbol{\omega}_{k}-\boldsymbol{n}_{3}\cdot\boldsymbol{\omega}_{3})},\label{Eq.Fourier_phi}
	\end{align}
\end{subequations}
where the summation $\sum_{\boldsymbol{n}_{k}=-\infty}^{\infty}$ is taken for all the combinations of number-vector components $(n_{k\alpha},n_{k\beta},n_{k\gamma})$. The conjugate $\left[A_{\boldsymbol{n}_{k}\boldsymbol{n}_{l}}\right]^{*}$ of the Fourier coefficient $A_{\boldsymbol{n}_{k}\boldsymbol{n}_{l}}$ has the property
\begin{align}
A_{-\boldsymbol{n}_{k}-\boldsymbol{n}_{l}}(\boldsymbol{I}_{k},\boldsymbol{I}_{l})=\left[A_{\boldsymbol{n}_{k}\boldsymbol{n}_{l}}(\boldsymbol{I}_{k},\boldsymbol{I}_{l})\right]^{*}.\label{Eq.conjugate_A}
\end{align}
Employing the Fourier transform of the pairwise potential, equation \eqref{Eq.Fourier_phi},  one can reduce equation \eqref{Eq.g_A} to
\begin{align}
g_{A}(1,2,t)=&-\frac{8i\pi^{3}}{N}\int^{t}_{t-\tau}\text{d}t'\int\text{d}\boldsymbol{I}_{3}\sum_{\boldsymbol{n}_{1}=-\infty}^{\infty}\sum_{m=1}^{2}\nonumber\\
&\times\left(A_{\boldsymbol{n}_{1}\boldsymbol{0}}\left(\boldsymbol{I}_{m}(t'),\boldsymbol{I}_{3}\right)e^{i\boldsymbol{n}_{1}\cdot\boldsymbol{\omega}_{m}t'}\boldsymbol{n}_{1}\cdot\frac{\partial}{\partial \boldsymbol{I}_{m}(t')}\right)\,  f\left(\boldsymbol{I}_{1}\left(t'\right),t'\right)\, f\left(\boldsymbol{I}_{2}\left(t'\right),t'\right)\, f\left(\boldsymbol{I}_{3},t\right),
\end{align}
where the Kronecker delta function is employed
\begin{align}
\delta_{\boldsymbol{n}_{3},\boldsymbol{0}}=\frac{1}{8\pi^{3}}\int\text{d}\boldsymbol{\omega}_{3}\, e^{-i\boldsymbol{n}_{3}\cdot\boldsymbol{\omega}_{3}}.\label{Eq_delta_fuc}
\end{align}
The Hamiltonian equation \eqref{Eq.Hamilton_eqns_Iw} reduces $g_{A}(1,2,t)$ to
\begin{align}
g_{A}(1,2,t)=&-\frac{8i\pi^{3}}{N}\int^{\tau}_{0}\text{d}\tau'\int\text{d}\boldsymbol{I}_{3}\sum_{\boldsymbol{n}_{1}=-\infty}^{\infty}\sum_{m=1}^{2}\nonumber\\
&\times\left(A_{\boldsymbol{n}_{1}\boldsymbol{0}}\left(\boldsymbol{I}_{m}\left(t-\tau'\right),\boldsymbol{I}_{3}\right)e^{i(\boldsymbol{n}_{1}\cdot\boldsymbol{\omega}_{m}-\boldsymbol{n}_{1}\cdot\Omega_{m} \tau')}\boldsymbol{n}_{1}\cdot\frac{\partial}{\partial \boldsymbol{I}_{m}(t-\tau')}\right)\nonumber\\
&\hspace{0.5cm}\times f\left(\boldsymbol{I}_{1}\left(t-\tau'\right),t-\tau'\right)\, f\left(\boldsymbol{I}_{2}\left(t-\tau'\right),t-\tau'\right)\, f\left(\boldsymbol{I}_{3},t\right).\label{Eq.g_A_fin}
\end{align}
Assuming that the Markovian approximation\footnote{`Markovian approximation' in kinetic theories is not the standard Markovian approximation to cut a series of random events for a two-point DF \citep[See, e.g.,][for Markovian approximations used for stellar dynamics.]{Kandrup_1980,Chavanis_2013a}. In kinetic theories, the approximation means that the (one-point) DF does not depend on the history of the system's physical characteristics or the previous motion of the test particle and field particles. Accordingly, $F(1(t-\tau),t-\tau)=F(1(t),t)$.} holds for the DFs and Fourier coefficients $A_{\boldsymbol{n}_{1}\boldsymbol{0}}$ and taking the limit $\tau \to \infty$, one obtains 
\begin{align}
g_{A}(1,2,t)=&\frac{8\pi^{3}}{N}\sum_{\boldsymbol{n}_{1}=-\infty}^{\infty}\sum_{m=1}^{2}\nonumber\\
&\times\left(\int\text{d}\boldsymbol{I}_{3}\, f(\boldsymbol{I}_{3},t)\, A_{\boldsymbol{n}_{1}\boldsymbol{0}}(\boldsymbol{I}_{m},\boldsymbol{I}_{3})\right)\frac{e^{i\boldsymbol{n}_{1}\cdot\boldsymbol{\omega}_{m}}}{-\boldsymbol{n}_{1}\cdot\Omega_{m}+i\epsilon}\, \boldsymbol{n}_{1}\cdot\frac{\partial}{\partial \boldsymbol{I}_{m}}\, f(\boldsymbol{I}_{1},t)\,f(\boldsymbol{I}_{2},t),\label{eq:gA_expression}
\end{align}
where the following identity is employed
\begin{align}
\int^{\infty}_{0}\text{d}t\,e^{iat}=\frac{1}{a+i\epsilon},
\end{align}
where $\epsilon$ is a diminishing positive constant and to be taken zero at the end of the formulation. After inserting equation \eqref{eq:gA_expression} into equation \eqref{Eq.collision_gLandau_orb} and executing a proper calculation (see Appendix \ref{Appendix_CA} for the detailed derivation), one can derive the explicit expression of the collision term $C_{A}$ due to $g_{A}(1,2,t)$ in action-angle coordinates 
\begin{align}
C_{A}\left[f(\boldsymbol{I}_{1},t)\right]=&-\frac{\pi(8\pi^{3})^{2}}{N}\sum^{\infty}_{\boldsymbol{n}_{1}=-\infty}\boldsymbol{n}_{1}\cdot\frac{\partial}{\partial \boldsymbol{I}_{1}}\nonumber\\
&\hspace{2cm}\times\left(\left|\int\text{d}\boldsymbol{I}_{2}A_{\boldsymbol{n}_{1}\boldsymbol{0}}(\boldsymbol{I}_{1},\boldsymbol{I}_{2})f(\boldsymbol{I}_{2},t)\right|^{2}\delta(\boldsymbol{n}_{1}\cdot\Omega_{1})\boldsymbol{n}_{1}\cdot\frac{\partial f(\boldsymbol{I}_{1},t)}{\partial \boldsymbol{I}_{1}}\right).\label{Eq.C_A_final}
\end{align}

\subsubsection{The explicit expression of the orbit-averaged g-Landau equation}\label{subsec:glanda_WI}

As discussed in \citep{Polyachenko_1982}, the collision term due to the correlation function $g_{a}(1,2,t)$ has the following form
\begin{align}
C_{a}(f(\boldsymbol{I}_{1},t))=&\pi(8\pi^{3})\sum^{\infty}_{\boldsymbol{n}_{1}=-\infty}\sum^{\infty}_{\boldsymbol{n}_{2}=-\infty}\int\text{d}\boldsymbol{I}_{2}\,\boldsymbol{n}_{1}\cdot\frac{\partial}{\partial \boldsymbol{I}_{1}}\nonumber\\
&\times\left(\left|A_{\boldsymbol{n}_{1}\boldsymbol{n}_{2}}(\boldsymbol{I}_{1},\boldsymbol{I}_{2})\right|^{2}\delta(\boldsymbol{n}_{1}\cdot\boldsymbol{\Omega}_{1}-\boldsymbol{n}_{2}\cdot\boldsymbol{\Omega}_{2})\left(\boldsymbol{n}_{1}\cdot\frac{\partial}{\partial \boldsymbol{I}_{1}}-\boldsymbol{n}_{2}\cdot\frac{\partial}{\partial \boldsymbol{I}_{2}}\right)f(\boldsymbol{I}_{1},t)\,f(\boldsymbol{I}_{2},t)\right),
\end{align}
The newly derived collision term $C_{A}(f(\boldsymbol{I}_{1},t))$ is negative, while we must characterize the relaxation process by the square of the fluctuation in the discreteness stochastic acceleration, equation \eqref{Eq.dis_fluc_order}. Hence,  we combine the two collision terms $C_{A}$ and $C_{a}$ in a more convenient form to extract the fluctuation effect. To do so, first, consider the collision terms associated with the many-body relaxation effect $(\boldsymbol{n}_{2}=\boldsymbol{0})$ only. (See Section \ref{sec:import_discrete} for the relationship of the many-body effect and $\boldsymbol{n}_{2}=\boldsymbol{0}$). Then, label the summation of the collision terms as $C_{a}^{(\boldsymbol{n}_{2}=0)}$;
\begin{align}
C_{a}^{(\boldsymbol{n}_{2}=0)}
=&-C_{A}+\-\frac{\pi(8\pi^{3})^{2}}{N}\sum^{\infty}_{\boldsymbol{n}_{1}=-\infty}\boldsymbol{n}_{1}\cdot\frac{\partial}{\partial \boldsymbol{I}_{1}}\delta(\boldsymbol{n}_{1}\cdot\boldsymbol{\Omega}_{1})\, \boldsymbol{n}_{1}\cdot\frac{\partial f(\boldsymbol{I}_{1},t)}{\partial \boldsymbol{I}_{1}}\nonumber\\
&\hspace{1.9cm}\times\left(N\int\text{d}\boldsymbol{I}_{2}\, f(\boldsymbol{I}_{2},t)\left|A_{\boldsymbol{n}_{1}\boldsymbol{0}}(\boldsymbol{I}_{1},\boldsymbol{I}_{2})\right|^{2}-\left|\int\text{d}\boldsymbol{I}_{2}\,A_{\boldsymbol{n}_{1}\boldsymbol{0}}(\boldsymbol{I}_{1},\boldsymbol{I}_{2})\,f(\boldsymbol{I}_{2},t)\right|^{2}\right).\label{Eq.Ca}
\end{align}
By replacing the total number $N$ in the bracket on the right-hand side of equation \eqref{Eq.Ca} by its integral representation, $N=(2\pi)^{3}\int\text{d}\boldsymbol{I}_{1}f(\boldsymbol{I}_{1},t)$, one has
\begin{align}
&C_{a}^{(\boldsymbol{n}_{2}=0)}+C_{A}\nonumber\\
&=\frac{\pi(8\pi^{3})^{2}}{2N}\sum^{\infty}_{\boldsymbol{n}_{1}=-\infty}\boldsymbol{n}_{1}\cdot\frac{\partial}{\partial \boldsymbol{I}_{1}}\nonumber\\
&\hspace{0.5cm}\times\left\{\iint\text{d}\boldsymbol{I}_{2}\,\text{d}\boldsymbol{I}_{3}\, f(\boldsymbol{I}_{2},t)\,f(\boldsymbol{I}_{3},t)\left|A_{\boldsymbol{n}_{1}\boldsymbol{0}}(\boldsymbol{I}_{1},\boldsymbol{I}_{2})-A_{\boldsymbol{n}_{1}\boldsymbol{0}}(\boldsymbol{I}_{1},\boldsymbol{I}_{3})\right|^{2}\delta(\boldsymbol{n}_{1}\cdot\boldsymbol{\Omega}_{1})\, \boldsymbol{n}_{1}\cdot\frac{\partial f(\boldsymbol{I}_{1},t)}{\partial \boldsymbol{I}_{1}}\right\},
\end{align}
where the following mathematical identity is employed
	\begin{align}
	&\int\text{d}\boldsymbol{I}_{2}\, \int\text{d}\boldsymbol{I}_{3}\, f(\boldsymbol{I}_{2},t)\, f(\boldsymbol{I}_{3},t)\, \left(\left|A_{\boldsymbol{n}_{1}\boldsymbol{0}}(\boldsymbol{I}_{1},\boldsymbol{I}_{2})\right|^{2}-A_{\boldsymbol{n}_{1}\boldsymbol{0}}(\boldsymbol{I}_{1},\boldsymbol{I}_{2})\left[A_{\boldsymbol{n}_{1}\boldsymbol{0}}(\boldsymbol{I}_{1},\boldsymbol{I}_{3})\right]^{*}\right)\nonumber\\
	&\hspace{3cm}=\frac{1}{2}\int\text{d}\boldsymbol{I}_{2}\int\text{d}\boldsymbol{I}_{3}\, f(\boldsymbol{I}_{2},t)\, f(\boldsymbol{I}_{3},t)\, \left|A_{\boldsymbol{n}_{1}\boldsymbol{0}}(\boldsymbol{I}_{1},\boldsymbol{I}_{2})-A_{\boldsymbol{n}_{1}\boldsymbol{0}}(\boldsymbol{I}_{1},\boldsymbol{I}_{3})\right|^{2}.
	\end{align}
Hence, including both the two-body relaxation effect $(\boldsymbol{n}_{2}\neq\boldsymbol{0})$ and many-body one $(\boldsymbol{n}_{2}=\boldsymbol{0})$, the orbit-averaged g-Landau kinetic equation reads
\begin{align}
&\partial_{t}f(\boldsymbol{I}_{1},t)\nonumber\\
&=\pi(8\pi^{3})\sum^{\infty}_{\boldsymbol{n}_{1}=-\infty}\sum^{\infty}_{\boldsymbol{n}_{2}=-\infty\&\boldsymbol{n}_{2}\neq0}\int\text{d}\boldsymbol{I}_{2}\,\boldsymbol{n}_{1}\cdot\frac{\partial}{\partial \boldsymbol{I}_{1}}\nonumber\\
&\hspace{0.9cm}\times\left(\left|A_{\boldsymbol{n}_{1}\boldsymbol{n}_{2}}(\boldsymbol{I}_{1},\boldsymbol{I}_{2})\right|^{2}\delta(\boldsymbol{n}_{1}\cdot\boldsymbol{\Omega}_{1}-\boldsymbol{n}_{2}\cdot\boldsymbol{\Omega}_{2})\left(\boldsymbol{n}_{1}\cdot\frac{\partial}{\partial \boldsymbol{I}_{1}}-\boldsymbol{n}_{2}\cdot\frac{\partial}{\partial \boldsymbol{I}_{2}}\right)f(\boldsymbol{I}_{1},t)\,f(\boldsymbol{I}_{2},t)\right)\nonumber\\
&\hspace{0.5cm}+\frac{\pi(8\pi^{3})^{2}}{2N}\sum^{\infty}_{\boldsymbol{n}_{1}=-\infty}\boldsymbol{n}_{1}\cdot\frac{\partial}{\partial \boldsymbol{I}_{1}}\nonumber\\
&\hspace{0.9cm}\times\left(\iint\text{d}\boldsymbol{I}_{2}\text{d}\boldsymbol{I}_{3}\left|A_{\boldsymbol{n}_{1}\boldsymbol{0}}(\boldsymbol{I}_{1},\boldsymbol{I}_{2})-A_{\boldsymbol{n}_{1}\boldsymbol{0}}(\boldsymbol{I}_{1},\boldsymbol{I}_{3})\right|^{2}\, f(\boldsymbol{I}_{2},t)\, f(\boldsymbol{I}_3,t)\, \delta(\boldsymbol{n}_{1}\cdot\boldsymbol{\Omega}_{1})\, \boldsymbol{n}_{1}\cdot\frac{\partial f(\boldsymbol{I}_{1},t)}{\partial \boldsymbol{I}_{1}}\right).\label{Eq.orb_gLandau}
\end{align}
In the standard stellar-dynamics theory, the relaxation process is dominated by the two-body relaxation with a local, homogeneous approximation. (See Appendix \ref{sec:conserv_action_integrals} for the relationship between the orbit-averaged g-Landau equation and local, homogeneous two-body relaxation in the FP equation.) The many-body relaxation is not a primary effect. Accordingly, the many-body effect $(\boldsymbol{n}_{2}=0)$ is negligible \citep[e.g.,][]{Polyachenko_1982}. The effect of $\boldsymbol{n}_{2}=0$ is possibly important to fine-tune the Coulomb logarithm and discuss the early relaxation-evolution stage of star clusters (Section \ref{sec:import_discrete}). For the latter, we consider only an ideal condition that stars less approach each other in physical spaces, and the cluster is a self-organized system approaching the state that resonant condition $\boldsymbol{n}_{1}\cdot\boldsymbol{\Omega}_{1}=0$ could happen.
\section{The basic properties of the orbit-averaged g-Landau equation}\label{sec:prop}

The stationary solution of the orbit-averaged g-Landau equation \eqref{Eq.orb_gLandau} is obviously the Maxwell-Boltzmann DF
\begin{align}
f(\boldsymbol{I}_{1},t)=\alpha e^{-\beta h(\boldsymbol{I}_{1})},
\end{align}
where $\alpha$ and $\beta$ are constant. To show other properties of the collision term in the orbit-averaged g-Landau equation \eqref{Eq.orb_gLandau}, we first rewrite the expression of collision term $C(f(\boldsymbol{I}_{1},t))$ as
\begin{subequations}
	\begin{align}
&C(f(\boldsymbol{I}_{1},t)) =\frac{\partial}{\partial\boldsymbol{I}_{1}}\cdot \boldsymbol{\mathcal{F}}(f(\boldsymbol{I}_{1},t)),\label{Eq_flux}\\
&\boldsymbol{\mathcal{F}}(f(\boldsymbol{I}_{1},t))\equiv\pi(8\pi^{3})\sum^{\infty}_{\boldsymbol{n}=-\infty}\sum^{\infty}_{\boldsymbol{n}_{1}=-\infty\&n_{1}\neq0}\boldsymbol{n} \int\text{d}\boldsymbol{I}_{2}\left|A_{\boldsymbol{nn_{1}}}(\boldsymbol{I}_{1},\boldsymbol{I}_{2})\right|^{2}\delta(\boldsymbol{n}\cdot\Omega_{1}-\boldsymbol{n}_{1}\cdot\Omega_{2})\nonumber\\
&\hspace{7.5cm}\times\left(\boldsymbol{n}\cdot\frac{\partial}{\partial \boldsymbol{I}_{1}}-\boldsymbol{n}_{1}\cdot\frac{\partial}{\partial \boldsymbol{I}_{2}}\right)f(\boldsymbol{I}_{1},t)\,f(\boldsymbol{I}_{2},t)\nonumber\\
&\hspace{2cm}+\frac{\pi(8\pi^{3})^{2}}{2N}\sum^{\infty}_{\boldsymbol{n}=-\infty}\boldsymbol{n}\int\text{d}\boldsymbol{I}_{2}\int\text{d}\boldsymbol{I}_{3}\left|A_{\boldsymbol{n}\boldsymbol{0}}(\boldsymbol{I}_{1},\boldsymbol{I}_{2})-A_{\boldsymbol{n}\boldsymbol{0}}(\boldsymbol{I}_{1},\boldsymbol{I}_{3})\right|^{2}\nonumber\\
&\hspace{7.5cm}\times f(\boldsymbol{I}_{2},t)\,f(\boldsymbol{I}_3,t)\,\delta(\boldsymbol{n}\cdot\Omega_{1})\,\boldsymbol{n}\cdot\frac{\partial f(\boldsymbol{I}_{1},t)}{\partial \boldsymbol{I}_{1}}.
\end{align}
\end{subequations}
where the functional $\boldsymbol{\mathcal{F}}(f(\boldsymbol{I}_{1},t)) $ means an `action flux' in terms of the stellar DF, i.e., a change in the DF through a hypersurface of action spaces per time because of the relaxation process. The boundary condition for the flux is
\begin{equation}
\boldsymbol{\mathcal{F}}(\boldsymbol{I}_\text{max})=0=\boldsymbol{\mathcal{F}}(\boldsymbol{I}_\text{min}).\label{Eq.Flux_BC}
\end{equation}
We discuss the anti-normalization, the conservation laws, and the \emph{H}-theorem in Sections \ref{subsec:anti}, \ref{subsec.conserv}, and \ref{subsec.H_theorem}, respectively.

\subsection{Anti-normalization condition}\label{subsec:anti}

As explained in Section \ref{sec:Importance_gLandau}, the anti-normalization condition is a fundamental property to define the DF, the total number, and the total energy of stars. The correlation function $g_{a}(1,2,t)$, equation \eqref{Eq.g_a(1,2,t)}, derived in \citep{Polyachenko_1982} does not satisfy the condition
\begin{align}
\int\text{d}2\, g_{a}(1,2,t)=\pi(8\pi^{3})\,i\,\sum_{\boldsymbol{n}=-\infty}^{\infty}\int\text{d}\boldsymbol{I}_{1}A_{\boldsymbol{n}\boldsymbol{0}}(\boldsymbol{I}_{1},\boldsymbol{I}_{2})e^{i\boldsymbol{n}\cdot\omega_{1}}\delta(\boldsymbol{n}\cdot\boldsymbol{\omega}_{1})\boldsymbol{n}\cdot\frac{\partial }{\partial \boldsymbol{I}_{1}}f(\boldsymbol{I}_{1})\,f(\boldsymbol{I}_{2})\neq0.\label{Eq.anti_ga}
\end{align}
On the one hand, the contribution from the correlation function $g_{A}(1,2,t)$ is
\begin{align}
\int\text{d}2\, g_{A}(1,2,t) =-\pi(8\pi^{3})\,i\,\sum_{\boldsymbol{n}=-\infty}^{\infty}\int\text{d}\boldsymbol{I}_{1}\, A_{\boldsymbol{n}\boldsymbol{0}}(\boldsymbol{I}_{1},\boldsymbol{I}_{2})\, e^{i\boldsymbol{n}\cdot\omega_{1}}\delta(\boldsymbol{n}\cdot\boldsymbol{\omega}_{1})\, \boldsymbol{n}\cdot\frac{\partial }{\partial \boldsymbol{I}_{1}}f(\boldsymbol{I}_{1})\,f(\boldsymbol{I}_{2}),\label{Eq.anti_gA}
\end{align}
where the integral representation for the Kronecker delta function, equation \eqref{Eq_delta_fuc}, is employed. Hence, the orbit-averaged g-Landau equation can satisfy the anti-normalization condition, equation \eqref{Eq.anti-normal},
\begin{align}
\int\text{d}2\, \left[\, g_{a}(1,2,t)+g_{A}(1,2,t)\, \right]=0.
\end{align}

That the orbit-averaged g-Landau equation holds the anti-normalization condition is important from the following points of view. First, one can correctly define the DF of stars in terms of action spaces with the orbit-averaging approximation, as explained in Section \ref{sec:basic_kinetics}. The failure of the anti-normalization condition prevents us from employing the (one-body) DF. For example, without the correlation function $g_{A}(1,2,t)$, one can not correctly define the DF in the BBGKY hierarchy
\begin{align}
\int\text{d}2\, \frac{f(1,2,t)}{N-1}&=\int\text{d}2 \left[\frac{f(1,t)\,f(2,t)}{N}+\frac{1}{N-1}g_{a}(1,2,t)\right]=f(1,t)+\int\text{d}2\,\frac{1}{N-1}\,g_{a}(1,2,t),\nonumber\\
&\neq f(1,t).
\end{align}  
Second, the action-angle approach can apply to various coordinates such as spherical, cylindrical, disk, and spheroidal systems. Considering this broad range of applicability, it is imperative for a kinetic equation in action-angle coordinates to satisfy a statistical theory's fundamental demands, like the anti-normalization condition.  Third, the anti-normalization condition allows us to correctly define the `discreteness' fluctuation in the standard pairwise acceleration, equation \eqref{Eq.dis_fluc_order}, in kinetic formulations under the orbit-averaging approximation. Without the anti-normalization condition, the BBGKY hierarchy only provides the fluctuation $ \left<\boldsymbol{a}_{12}(t)\,\tilde{\boldsymbol{a}}_{12}\left(t'\right)\right>_{\boldsymbol{A}\neq 0}$ but not $ \left<\tilde{\boldsymbol{a}}_{12}(t)\,\tilde{\boldsymbol{a}}_{12}\left(t'\right)\right>_{\boldsymbol{A}\neq 0}$;\footnote{It appears that the factor $ \frac{\boldsymbol{A}_{1}}{N}\cdot\boldsymbol{\partial}_{1}\int g_{a}(1,2,t;\tilde{\boldsymbol{a}}_{12})\,\text{d}2$ has subtle importance for practical applications. However, the kinetic formulation must include the factor when considering the many-body relaxation strictly based on the projected-operator theory \citep{Kandrup_1981}.}
\begin{subequations}
	 \begin{align}
	&\ll\tilde{\boldsymbol{a}}_{12}(t)\,\tilde{\boldsymbol{a}}_{12}\left(t'\right)\gg_{\boldsymbol{A}\neq 0} \,\,\,\neq\,\,\, \ll\boldsymbol{a}_{12}(t)\,\tilde{\boldsymbol{a}}_{12}\left(t'\right)\gg_{\boldsymbol{A}\neq 0},\\
	<=>&\int\text{d}t'\iint\text{d}2(t)\,\text{d}2\left(t'\right)\,\tilde{\boldsymbol{a}}_{12}(t)\,\tilde{\boldsymbol{a}}_{12}(t')\,f(2(t),t)f\left(2\left(t'\right),t'\right)\nonumber\\
	&\hspace{3.6cm}\neq\int\text{d}t'\iint\text{d}2(t)\,\text{d}2\left(t'\right)\,\boldsymbol{a}_{12}(t)\,\tilde{\boldsymbol{a}}_{12}(t')\,f(2(t),t)f\left(2\left(t'\right),t'\right),\\
	<=>&\int\text{d}2\, \tilde{\boldsymbol{a}}_{12}\,g_{a}(1,2,t;\tilde{\boldsymbol{a}}_{12})\neq\int\text{d}2\, \boldsymbol{a}_{12}\,g_{a}(1,2,t;\tilde{\boldsymbol{a}}_{12}),
	\end{align}\label{Eq.dsic_fluc}
\end{subequations}
where the last line is the corresponding integral representation of the discrete fluctuation.

\subsection{Conservation of the total energy and the total number of stars}\label{subsec.conserv}

The anti-normalization condition holds for the orbit-averaged g-Landau equation. Hence, the total number of stars, equation \eqref{Def.N2_E2}, is conserved as follows
\begin{equation}
\frac{\text{d}N(t)}{\text{d}t}=8\pi^{3}\int\text{d}\boldsymbol{I}_{1}\, \frac{\partial f(\boldsymbol{I}_{1},t)}{\partial t}=8\pi^{3}\int\text{d}\boldsymbol{I}_{1}\, \frac{\partial}{\partial\boldsymbol{I}_{1}}\,\boldsymbol{\mathcal{F}}(f(\boldsymbol{I}_{1},t))=0,\label{Eq.consv_N}
\end{equation}
where the g-Landau equation \eqref{Eq.orb_gLandau} and equation \eqref{Eq_flux} are employed for the second equality. The boundary condition, equation \eqref{Eq.Flux_BC}, is also employed for the third equality. We define the sum of kinetic energy and (quasi-)static potential energy as 
\begin{align}
E_\text{st}(t)=\int\text{d}1\, f(1,t)\,h(\boldsymbol{I}_{1})=8\pi^{3}\int\text{d}\boldsymbol{I}_{1}\, f(\boldsymbol{I}_{1},t)\,h(\boldsymbol{I}_{1}).
\end{align}
The energy can be conserved for the g-Landau equation as follows
\begin{subequations}
\begin{align}
\frac{\text{d}E_\text{st}(t)}{\text{d}t}=&-8\pi^{3}\int\text{d}\boldsymbol{I}_{1}\, \boldsymbol{\mathcal{F}}(f(\boldsymbol{I}_{1},t))\cdot\boldsymbol{\Omega}_{1},\\ =&-\frac{\pi(8\pi^{3})^{2}}{2}\sum^{\infty}_{\boldsymbol{n}=-\infty}\sum^{\infty}_{\boldsymbol{n}_{1}=-\infty\&n_{1}\neq0}\int\text{d}\boldsymbol{I}_{1}\int\text{d}\boldsymbol{I}_{2}\left(\boldsymbol{n}\cdot\boldsymbol{\Omega}_{1}-\boldsymbol{n}_{1}\cdot\boldsymbol{\Omega}_{2}\right)\delta(\boldsymbol{n}\cdot\boldsymbol{\Omega}_{1}-\boldsymbol{n}_{1}\cdot\boldsymbol{\Omega}_{2})\nonumber\\ 
&\hspace{3.5cm}\times\left\{\left|A_{\boldsymbol{n}\boldsymbol{n_{1}}}(\boldsymbol{I}_{1},\boldsymbol{I}_{2})\right|^{2}\left(\boldsymbol{n}\cdot\frac{\partial}{\partial \boldsymbol{I}_{1}}-\boldsymbol{n}_{1}\cdot\frac{\partial}{\partial \boldsymbol{I}_{2}}\right)f(\boldsymbol{I}_{1},t)f(\boldsymbol{I}_{2},t)\right\}\nonumber\\
&-\frac{\pi(8\pi^{3})^{3}}{2N}\sum^{\infty}_{\boldsymbol{n}=-\infty}\int\text{d}\boldsymbol{I}_{I}\int\text{d}\boldsymbol{I}_{2}\int\text{d}\boldsymbol{I}_{3}\,\boldsymbol{n}\cdot\boldsymbol{\Omega}_{1}\,\delta(\boldsymbol{n}\cdot\boldsymbol{\Omega}_{1})\nonumber\\
&\hspace{3.5cm}\times\left(\left|A_{\boldsymbol{n}\boldsymbol{0}}(\boldsymbol{I}_{1},\boldsymbol{I}_{2})-A_{\boldsymbol{n}\boldsymbol{0}}(\boldsymbol{I}_{1},\boldsymbol{I}_{3})\right|^{2}f(\boldsymbol{I}_{2},t)f(\boldsymbol{I}_3,t)\boldsymbol{n}\cdot\frac{\partial f(\boldsymbol{I}_{1},t)}{\partial \boldsymbol{I}_{1}}\right),\\
=&0,
\end{align}\label{Eq.consv_Est}
\end{subequations}
where the orbit-averaged g-Landau equation \eqref{Eq.orb_gLandau} and the Hamiltonian-Jacobi equation \eqref{Eq.Hamilton_Jacobi_eqns} are employed for the first equality. The Kronecker delta function, equation \eqref{Eq_delta_fuc}, is also used for the second equality.

The conservation of total stellar number is an obvious demand from the physical principle even at the orbit-averaged-equation level so long as the anti-normalization condition holds. On the one hand, the conservation of energy $E_\text{st}$ is not based on a fundamental principle. The conservation rule is merely a result of the artificial assumption that the integral vectors $\boldsymbol{I}_{1}$ do not change with time in relaxation processes.  Accordingly, one-body Hamiltonian (equation \eqref{Eq.one_body_Hamilton}) or the energy available to stars, is constant. Hence, the MF potential is independent of the test star's motion, and one loses a self-consistent formulation of the total energy $E(t)$. This assumption is necessary only for employing the action-angle-variables approach. However, the conservation of the energy $E_\text{st}$ has been useful to test results in numerically solving the local, homogeneous orbit-averaged FP equations \citep[e.g.,][]{Cohn_1979,Takahashi_1995}.

\subsection{$H$-theorem}\label{subsec.H_theorem}
We examine the $H$-theorem for the Boltzmann entropy. The $H$-function reads
\begin{align}
H(t)\equiv -\int\text{d}1\, f(1,t)\ln[f(1,t)].
\end{align}
One can prove that the $H$-function increases with time as follows
\begin{subequations}
\begin{align}
\frac{\text{d}H(t)}{\text{d}t}=&8\pi^{3}\int\text{d}\boldsymbol{I}_{1}\, \boldsymbol{\mathcal{F}}(f(\boldsymbol{I}_{1},t))\cdot\frac{\partial f(\boldsymbol{I}_{1})}{\partial \boldsymbol{I}_{1}}\frac{1}{f(\boldsymbol{I}_{1},t)},\\
     =&\frac{\pi(8\pi^{3})^{2}}{2}\sum^{\infty}_{\boldsymbol{n}=-\infty}\sum^{\infty}_{\boldsymbol{n}_{1}=-\infty\&\boldsymbol{n}_{1}\neq0}\iint\text{d}\boldsymbol{I}_{1}\text{d}\boldsymbol{I}_{2}\frac{\left|A_{\boldsymbol{nn_{1}}}(\boldsymbol{I}_{1},\boldsymbol{I}_{2})\right|^{2}}{f(\boldsymbol{I}_{1},t)\,f(\boldsymbol{I}_{2},t)}\delta(\boldsymbol{n}\cdot\boldsymbol{\Omega}_{1}-\boldsymbol{n}_{1}\cdot\boldsymbol{\Omega}_{2})\nonumber\\
&\hspace{5cm}\times\left[\left(\boldsymbol{n}\cdot\frac{\partial}{\partial \boldsymbol{I}_{1}}-\boldsymbol{n}_{1}\cdot\frac{\partial}{\partial \boldsymbol{I}_{2}}\right)f(\boldsymbol{I}_{1},t)f(\boldsymbol{I}_{2},t)\right]^{2}\nonumber\\
    &+\frac{\pi(8\pi^{3})^{3}}{2N}\sum^{\infty}_{\boldsymbol{n}=-\infty}\iiint\text{d}\boldsymbol{I}_{1}\,\text{d}\boldsymbol{I}_{2}\,\text{d}\boldsymbol{I}_{3}\frac{\left|A_{\boldsymbol{n}\boldsymbol{0}}(\boldsymbol{I}_{1},\boldsymbol{I}_{2})-A_{\boldsymbol{n}\boldsymbol{0}}(\boldsymbol{I}_{1},\boldsymbol{I}_{3})\right|^{2}}{f(\boldsymbol{I}_{1},t)\,f(\boldsymbol{I}_{2},t)\,f(\boldsymbol{I}_3,t)}\delta(\boldsymbol{n}\cdot\boldsymbol{\Omega}_{1})\nonumber\\
&\hspace{3.5cm}\times\left[\boldsymbol{n}\cdot\frac{\partial }{\partial \boldsymbol{I}_{1}}f(\boldsymbol{I}_{1},t)\,f(\boldsymbol{I}_{2},t)\,f(\boldsymbol{I}_3,t)\right]^{2},\\
\geq& 0,
\end{align}\label{Eq.Htheore}
\end{subequations}
where the orbit-averaged g-Landau equation \eqref{Eq.orb_gLandau} and the Hamiltonian-Jacobi equation \eqref{Eq.Hamilton_Jacobi_eqns} are employed for the second equality.

\section{Discussion: the discreteness stochastic acceleration in the relaxation evolution of star clusters}\label{sec:import_discrete}

The present section highlights the difference between the discreteness stochastic acceleration and two-body relaxation. It also discusses the acceleration's physical feature based on the orbit-averaged g-Landau equation \eqref{Eq.orb_gLandau}. (See Appendix \ref{sec:fourier_coeff} for the mathematical structure of the equation.) To exclude trivial factors, we rewrite the equation as follows
\begin{subequations}
\begin{align}
&\partial_{t}f(\boldsymbol{I}_{1},t)=\pi(8\pi^{3})\sum^{\infty}_{\boldsymbol{n}_{1}=-\infty}\boldsymbol{n}_{1}\cdot\frac{\partial}{\partial \boldsymbol{I}_{1}}\left(\sum^{\infty}_{\boldsymbol{n}_{2}=-\infty\&\boldsymbol{n}_{2}\neq0}\Tilde{F}_\text{two}(\boldsymbol{I}_{1},t;\boldsymbol{n}_{1},\boldsymbol{n}_{2})+\frac{8\pi^{3}}{2N}\Tilde{F}_\text{s.a.}(\boldsymbol{I}_{1},t;\boldsymbol{n}_{1})\right),\label{Eq.orb_gLandau_many}\\
&\Tilde{F}_\text{two}\equiv\int\text{d}\boldsymbol{I}_{2}\left(\left|A_{\boldsymbol{n}_{1}\boldsymbol{n}_{2}}(\boldsymbol{I}_{1},\boldsymbol{I}_{2})\right|^{2}\delta(\boldsymbol{n}_{1}\cdot\boldsymbol{\Omega}_{1}-\boldsymbol{n}_{2}\cdot\boldsymbol{\Omega}_{2})\left(\boldsymbol{n}_{1}\cdot\frac{\partial}{\partial \boldsymbol{I}_{1}}-\boldsymbol{n}_{2}\cdot\frac{\partial}{\partial \boldsymbol{I}_{2}}\right)f(\boldsymbol{I}_{1},t)f(\boldsymbol{I}_{2},t)\right),\\
&\Tilde{F}_\text{s.a.}\equiv\iint\text{d}\boldsymbol{I}_{2}\,\text{d}\boldsymbol{I}_{3}\left|A_{\boldsymbol{n}_{1}\boldsymbol{0}}(\boldsymbol{I}_{1},\boldsymbol{I}_{2})-A_{\boldsymbol{n}_{1}\boldsymbol{0}}(\boldsymbol{I}_{1},\boldsymbol{I}_{3})\right|^{2}\, f(\boldsymbol{I}_{2},t)\, f(\boldsymbol{I}_3,t)\, \delta(\boldsymbol{n}_{1}\cdot\boldsymbol{\Omega}_{1})\, \boldsymbol{n}_{1}\cdot\frac{\partial f(\boldsymbol{I}_{1},t)}{\partial \boldsymbol{I}_{1}}.
\end{align}
\end{subequations}
The factors $\{\Tilde{F}_\text{two}\}$ describe the two-body relaxation with inhomogeneous field stars, while $\{\Tilde{F}_\text{s.a.}\}$ represent the many-body relaxation because of the discreteness stochastic acceleration. This difference between the factors $\{\Tilde{F}_\text{two}\}$ and $\{\Tilde{F}_\text{s.a.}\}$ may be clear. In the standard kinetic formulations, many-body relaxation's effect is represented by the second field star or star 3 at the kinetic-equation level.\footnote{For example, in the standard plasma kinetic theory \citep[e.g.,][]{Balescu_1997,Liboff_2003}, stars 1 and 2 are considered `naked' and undergo the two-body relaxation process through the Coulomb potential if any other stars do not affect the relaxation process. However, in the long-time limit, another field star (star 3) comes into play. It takes the role of the response from the rest of the field stars perturbed by the test star. This perturbation adjusts the field particles' configuration (plasma polarization) and mathematically relies on the correlation functions $g(1,3,t)$ and $g(2,3,t)$ in the BBGKY hierarchy, like the gravitational polarization shown in equation \eqref{Eq.Cc}. The functions originate from the ternary DF $f(1,2,3,t)$. Because of the polarization effect from star 3, stars 1 and 2 can be `dressed' and interact through the Yukawa potential.} Hence, one may consider $\{\Tilde{F}_\text{s.a.}\}$ to represent the effect of many-body relaxation, or discreteness stochastic acceleration, through the DF $f(\boldsymbol{I}_{3},t)$. On the one hand, $\{\Tilde{F}_\text{two}\}$ describe only the two-body relaxation with inhomogeneous field stars. Accordingly, in the orbit-averaged g-Landau equation \eqref{Eq.orb_gLandau}, the condition  $\boldsymbol{n}_{2}=\boldsymbol{0}$ is corresponding to the many-body relaxation, $\{\Tilde{F}_\text{s.a.}\}$, and $\boldsymbol{n}_{2}\neq\boldsymbol{0}$ to the two-body relaxation, $\{\Tilde{F}_\text{two}\}$. The discreteness stochastic acceleration may also be interpreted as the result of the orbit-averaged Newtonian-potential interaction between the test star and field stars (Appendix \ref{sec:fourier_coeff}), while the two-body relaxation is because of the `naked' Newtonian pairwise potential.  

The factors $\{\Tilde{F}_\text{two}\}$ and $\{\Tilde{F}_\text{s.a.}\}$ show an  independent property at the kinetic-equation level, as shown in Section \ref{sec:prop}. The factors independently hold the anti-normalization condition, the conservation laws of number $N$ and energy $E_\text{st}$, and the $H$-theorem, as seen from the mathematical expressions in equations \eqref{Eq.anti_ga} and \eqref{Eq.anti_gA}, equations  \eqref{Eq.consv_N} and \eqref{Eq.consv_Est}, and equation \eqref{Eq.Htheore}, respectively. In other words,  $\{\Tilde{F}_\text{s.a.}\}$ do not rely on $\{\Tilde{F}_\text{two}\}$ to form these properties. Hence, we can infer that the orbit-averaging process differentiates stars undergoing the many-body relaxation from those experiencing the two-body one. This independence is a clear advantage to discuss the early stage of the relaxation evolution since one would like to single out the effect of the many-body relaxation from the entire relaxation process, as the previous works discussed \citep{Kandrup_1981,Kandrup_1985,Ito_2020_3}.

Unlike the two-body relaxation process, the Boltzmann-Maxwellian DF is not only the stationary solution to the orbit-averaged g-Landau equation at the early stage of the relaxation evolution. Assume that stellar distances are so large that the many-body relaxation dominates the evolution and are mathematically described by the ideal condition $r_{12}\sim R$, following the discussion in \citep{Kandrup_1981}. Then, one may reduce equation \eqref{Eq.orb_gLandau_many} to
\begin{align}
&\partial_{t}f(\boldsymbol{I}_{1},t)=\frac{\pi(8\pi^{3})^{2}}{2N}\sum^{\infty}_{\boldsymbol{n}_{1}=-\infty}\boldsymbol{n}_{1}\cdot\frac{\partial}{\partial \boldsymbol{I}_{1}}\Tilde{F}_\text{s.a.}.\label{Eq.g-Landau_early}
\end{align}
This reduction appears based on a strong assumption. However, as pointed out in Section \ref{sec:BBGHY_action_angle}, we believe that it is proper to treat an orbit-averaged kinetic equation with the many-body relaxation only, considering the nature of the orbit-averaging and the periodicity of stellar orbits. A stationary solution to equation \eqref{Eq.g-Landau_early} is any  function $A$ of the one-body Hamiltonian
\begin{align}
    f(\boldsymbol{I}_{1},t)=A[h(\boldsymbol{I}_{1})].\label{Eq.Soln_g-Landau_early}
\end{align}
Hence, the cluster can reach various isotropic states besides the Maxwellian DF. Although we must analyze the stability of the solution for practical applications, the solution appears consistent with the previous works \citep{Kandrup_1985,Taruya_2003_3,Ito_2020_3}. The works showed that clusters could have a non-local-Maxwellian DF at the early stage of the relaxation evolution, especially a polytropic DF, $f(\boldsymbol{I}_{1},t)=\left[h(\boldsymbol{I}_{1})\right]^{m}$, where $m$ is the polytropic index. The polytropic structures appear in the observed structural profiles of low-concentration globular clusters with relatively long relaxation times \citep{Ito_2020_3}. The low-concentration clusters are still in their initial secular-evolution time. We may observe `isotropized' initial information of the star clusters before the two-body relaxation process deletes it. Accordingly, their current structures  may provide hints of the origin of star clusters.

The discreteness stochastic acceleration is different from collisionless relaxation (the violent relaxation, phase mixing, and chaotic mixing \citep[e.g.,][]{Saslaw_1985, Binney_2011}). The latter can drive a star cluster into a quasi-stationary state caused by a rapid fluctuation in the MF potential of a collisionless self-gravitating system. Because of the strong Jeans theorem, the corresponding stellar DF can be written in terms of  isolating integrals, like equation \eqref{Eq.Soln_g-Landau_early}. However, the collisionless relaxation is apparent and does not change the value of the  $H$-function with time. It does not provide a clear `direction' of the dynamical evolution, unlike the discreteness stochastic acceleration. The latter further limits the possible stationary DF so that the DF is isotropized.

The fact that the discreteness stochastic acceleration can isotropize the stellar DF appears sensible if we analogically consider cosmic rays’ statistical acceleration caused by fluctuating magnetic fields. Recall the mathematical form of the discreteness stochastic acceleration, equations \eqref{Eq.dis_fluc_null} and \eqref{Eq.dis_fluc_order}. The equations describe a similar condition to the fluctuating magnetic field that causes the stochastic acceleration, or the second-order Fermi acceleration \citep{Fermi_1949}, of charged particles. Cosmic rays are cumulatively scattered because of random magnetic fluctuations while propagating in interstellar mediums, interplanetary space, or  magnetized turbulent plasmas from their sources \citep[e.g.,][]{Melrose_2009, Shalchi_2009, Petrosian_2012}. The fundamental physical features in the relaxation process of cosmic rays appear similar to the two-body relaxation process in star clusters. Both the systems are inhomogeneous and undergo velocity relaxations because of resonant phenomena. A resonance occurs between charged particles and waves in cosmic rays while between two stellar orbits in star clusters. These similarities may be reasonable since both the scattering of cosmic rays \citep{Sturrock_1966, Jokipii_1966} and two-body relaxation of stars \citep{Rosenbluth_1957,Henon_1961} have been considered random-walk processes and mathematically modeled by FP-type kinetic equations proposed in \citep{Chandra_1943c}. The statistical acceleration of cosmic rays suggests that thermalization- and isotropization- times do not need to be in the same order. In a scattering-dominant magnetic turbulent system, the cosmic rays can be isotropized due to resonant (pitch-angle) scattering before they are relaxed to the thermal equilibrium \citep[e.g.,][]{Melrose_2009,Shalchi_2009}. In fact, a simple stationary FP model shows a power-law profile of cosmic-ray energy flux that matches an observed energy spectrum  \citep{De_Vega_2003}. Also, the isotropization of cosmic rays is not limited to collisional systems. It may occur even in collisionless plasma systems where the guiding centers of the gyromotion of charges jump from their original magnetic-field lines to  new ones, obeying a random walk \citep{Newman_2020}. 

However, the discreteness stochastic acceleration does not fall into the  statistical acceleration of cosmic rays and may apply to other self-gravitating systems.  We consider the former acceleration to be a new kind of stochastic acceleration caused by many-body relaxation. It increases the $H$-function but does not thermalize stars. Perhaps, the most relevant stellar-system model to the present system is a test particle in a  host collisionless system's Newtonian MF field, perturbed by the stochastic gravitational field from extended substructures \citep{Pe_arrubia_2017}.  Regardless of a clear difference in the time scale, test-particle method, and averaging process from the present system, \cite{Pe_arrubia_2019} showed that the tidal stochastic heating from the extended substructures isotropized the particle DF in a Keplerian potential. Because of the similarity between the fluctuating gravitational-field sources of the two systems, comparing the systems would be intriguing to clarify the nature of the discrete stochastic acceleration. Also, it can provide the possibility of applying the acceleration to other self-gravitating systems, such as galaxies and dark matter.

\section{Conclusion}\label{sec:conclusion}The most fundamental kinetic equation for finite weakly-coupled star clusters is \citep{Gilbert_1968}'s kinetic equation. Its mathematical expression is complicated. Hence, one has derived the approximates of the kinetic equation. However, only the g-Landau kinetic equation correctly approximates the \citep{Gilbert_1968}'s equation since it  satisfies basic demands from fundamental statistical mechanics, unlike the other star-cluster kinetic equations (Section \ref{sec:Importance_gLandau}). The present paper aimed to deepen our understanding of the g-Landau equation and the discreteness stochastic acceleration. 

We first derived the orbit-averaged g-Landau equation \eqref{Eq.orb_gLandau} in action-angle variables by extending  \citep{Polyachenko_1982}'s work to find the explicit form of the g-Landau equation (Section \ref{sec:explicit_orb_g_landau}).  The equation can conserve both the total stellar number and the sum of total stellar kinetic energy and quasi-static MF potential energy (Section \ref{sec:prop}). Also, it holds the $H$-theorem and anti-normalization condition as well. Section \ref{sec:import_discrete} discussed the property of the discreteness stochastic acceleration based on the orbit-averaged g-Landau equation. Stars undergoing the discreteness stochastic acceleration can satisfy the anti-normalization condition, the conservation laws, and the $H$-theorem, independently of stars experiencing the two-body relaxation. This feature helps us consistently treat only the discreteness stochastic acceleration of stars in the early stage of the relaxation evolution, neglecting the two-body relaxation. Then, we concluded that star clusters could reach an isotropic stationary state, possibly described by the polytrope models, other than the state described by the Maxwellian DF in the early relaxation-evolution stage. The isotropization of star clusters due to the discreteness statistical acceleration is a new possible observable feature of the clusters. The feature provides a different outcome from the two-body relaxation and the collisionless one.

In our future work, we will work on the following three direct extensions of the present work. (i) We will discuss the stability problem of the stationary solution, equation \eqref{Eq.Soln_g-Landau_early}, to the orbit-averaged g-Landau equation. (ii) We will combine the gravitational-polarization effect and the discreteness stochastic acceleration to discuss the stationary solution. This is since low-concentration globular clusters have shown the trait of primordial mass segregation \citep{De_Marchi_2007,Baumgardt_2008}. The gravitational polarization is expected to be significant in mass segregation \citep{Gilbert_1970}. (iii) We will specify the orbit-averaged g-Landau model's geometrical shape, e.g., spherical, disk-like, spheroidal, cylindrical, and asymmetric systems, with a proper approximation of the third integral. We will clarify the relationship of the discreteness stochastic acceleration with both the coordinate dependence and dimensionality at the orbit-averaged level.

\begin{acknowledgements}

\end{acknowledgements}

\appendix                  
\begin{appendices}
\renewcommand{\theequation}{\Alph{section}.\arabic{equation}}

\section{Anti-normalization for the homogeneous, local FP kinetic equation}\label{Appendix:anti}

The present appendix shows that the homogeneous, local FP kinetic equation can explicitly satisfy the anti-normalization condition. Assume that the stellar DF is local or homogeneous in equation \eqref{Eq.2ndBBGKY_Gilbert} 
\begin{align}
f(1,t)\,f(2,t)=\begin{cases}
                  n_\text{o}^{2}\, \chi(\boldsymbol{v}_{1},t)\, \chi(\boldsymbol{v}_{2},t),\qquad\text{(homogeneous)}\\
                  f(\boldsymbol{r}_{1},\boldsymbol{v}_{1},t)\, f(\boldsymbol{r}_{1},\boldsymbol{v}_{2},t),\qquad\text{(local)}
                 \end{cases}
\end{align}
where $\chi$ is any function of the momentum vector and time, and $n_\text{o}$ is the system density. Such a system approximates stellar encounters in the two-body relaxation evolution \citep{Henon_1965,Chavanis_2013a}. One can easily confirm that the integral of the FP equation \eqref{Eq.Cc} over the phase-space volume for star 2 vanishes as follows
\begin{align}
\int\text{d}2\, C_{c}\, =\left[\int\text{d}\boldsymbol{r}_{2}\,\boldsymbol{a}_{12}\right]\int\text{d}\boldsymbol{v}_{2}\, f(1,t)\,f(2,t)=0.
\end{align}
This feature relies on the local-encounter approximation that one can employ the mathematical approximation $\text{d}\boldsymbol{r}_{2}\approx\text{d}\boldsymbol{r}_{12}$. In fact, the explicit form of the correlation function for the FP collision term with homogeneous or local approximation satisfies the anti-normalization condition as follows 
\begin{align}
\int\text{d}2\, g(1,2,t)\propto\int^{\infty}_{0}\text{d}\tau\int\text{d}\boldsymbol{r}_{2}\, \boldsymbol{a}_{12}(t-\tau)=\int\text{d}\boldsymbol{r}_{12}\int^{\infty}_{0}\text{d}k\int^{\infty}_{0}\text{d}\tau\frac{Gmk\hat{k}}{2\pi^{2}i}e^{-i\boldsymbol{k}\cdot(\boldsymbol{r}_{12}-\boldsymbol{v}_{12}\tau)}=0,
\end{align}
where the unperturbed trajectory of the relative motion of stars follows the rectilinear motion. Also, the acceleration of star 1 due to the pair-wise Newtonian potential force from star 2 is Fourier-transformed \citep[see][for the explicit form of the FP equation.]{Kandrup_1981,Chavanis_2013a}

\section{Collision term $C_{A}$ in action-angle coordinates}\label{Appendix_CA}

In the present appendix, we derive the collision term $C_{A}$, equation \eqref{Eq.C_A_final}, in action-angle coordinates from the correlation function, equation \eqref{eq:gA_expression}. According to the angle-vector dependence, one may separate the correlation function $g_{A}(1,2,t)$ into the following functions
\begin{align}
&g_{A}(1,2,t)\equiv g_{A1}(\boldsymbol{I}_{1},\boldsymbol{I}_{2},\boldsymbol{\omega}_{1},t)+g_{A2}(\boldsymbol{I}_{1},\boldsymbol{I}_{2},\boldsymbol{\omega}_{2},t).
\end{align}
Then, one can systematically reduce the collision term $C(f(\boldsymbol{I}_{1},t))$, equation \eqref{Eq.collision_gLandau_orb}, to the explicit forms. The mathematical expression of each collision term is lengthy. Hence, we separate the collision terms due to the correlation functions $g_{A1}(\boldsymbol{I}_{1},\boldsymbol{I}_{2},\boldsymbol{\omega}_{1},t)$ and $g_{A2}(\boldsymbol{I}_{1},\boldsymbol{I}_{2},\boldsymbol{\omega}_{1},t)$ into four terms
\begin{subequations}
	\begin{align}
	&C_{A}\equiv C_{A1}+C_{A2},\label{Eq.collision}\\
	&C_{A1}\equiv C_{A11}+C_{A12},\label{Eq.collision_A1}\\
	&C_{A2}\equiv C_{A21}+C_{A22},\label{Eq.collision_A2}\\
	&C_{A11}\equiv\iint\frac{\text{d}2\,\text{d}\boldsymbol{\omega}_{1}}{8\pi^{3}}\left(\frac{\partial \phi_{12}}{\partial \boldsymbol{\omega}_{1}}\cdot\frac{\partial }{\partial \boldsymbol{I}_{1}}\right)g_{A1}(\boldsymbol{I}_{1},\boldsymbol{I}_{2},\boldsymbol{\omega}_{1},t),\label{Eq.collision_A11}\\
	&C_{A12}\equiv-\iint\frac{\text{d}2\,\text{d}\boldsymbol{\omega}_{1}}{8\pi^{3}}\left(\frac{\partial \phi_{12}}{\partial \boldsymbol{I}_{1}}\cdot\frac{\partial }{\partial \boldsymbol{\omega}_{1}}\right)g_{A1}(\boldsymbol{I}_{1},\boldsymbol{I}_{2},\boldsymbol{\omega}_{1},t),\label{Eq.collision_A12}\\
	&C_{A21}\equiv\iint\frac{\text{d}2\,\text{d}\boldsymbol{\omega}_{1}}{8\pi^{3}}\left(\frac{\partial \phi_{12}}{\partial \boldsymbol{\omega}_{1}}\cdot\frac{\partial }{\partial \boldsymbol{I}_{1}}\right)g_{A2}(\boldsymbol{I}_{1},\boldsymbol{I}_{2},\boldsymbol{\omega}_{2},t),\label{Eq.collision_A21}\\
	&C_{A22}\equiv-\iint\frac{\text{d}2\,\text{d}\boldsymbol{\omega}_{1}}{8\pi^{3}}\left(\frac{\partial \phi_{12}}{\partial \boldsymbol{I}_{1}}\cdot\frac{\partial }{\partial \boldsymbol{\omega}_{1}}\right)g_{A2}(\boldsymbol{I}_{1},\boldsymbol{I}_{2},\boldsymbol{\omega}_{2},t)=0.\label{Eq.collision_A22}
	\end{align}\label{Eq.collision_A1_gLandau_orb}
\end{subequations}
Employing equations \eqref{Eq.inv_Fourier_phi}  and \eqref{eq:gA_expression}, equation \eqref{Eq.collision_A11} reduces to
\begin{align}
C_{A11}=&-\frac{i}{N}\sum^{\infty}_{\boldsymbol{n}_{1}=-\infty}\sum^{\infty}_{\boldsymbol{n}_{2}=-\infty}\sum^{\infty}_{\boldsymbol{n}=-\infty}\iiint\text{d}2\,\text{d}\boldsymbol{\omega}_{1}\,\text{d}\boldsymbol{I}_{3}\,A_{\boldsymbol{n}_{1}\boldsymbol{n}_{2}}(\boldsymbol{I}_{1},\boldsymbol{I}_{2})\,e^{i([\boldsymbol{n}_{1}+\boldsymbol{n}]\cdot\boldsymbol{\omega}_{1}-\boldsymbol{n}_{2}\cdot\boldsymbol{\omega}_{2})}\nonumber\\
&\hspace{3cm}\times\boldsymbol{n}_{1}\cdot\frac{\partial}{\partial \boldsymbol{I}_{1}}\left\{\left[\frac{A_{\boldsymbol{n}\boldsymbol{0}}(\boldsymbol{I}_{1},\boldsymbol{I}_{3})}{-\boldsymbol{n}\cdot\boldsymbol{\Omega}_{1}+i\epsilon}\boldsymbol{n}\cdot\frac{\partial}{\partial \boldsymbol{I}_{1}}\right]f(\boldsymbol{I}_{1})\,f(\boldsymbol{I}_{2})\,f(\boldsymbol{I}_{3})\right\}.
\end{align}
Using the integral representation of Kronecker's delta function, equation \eqref{Eq_delta_fuc}, one obtains 
\begin{align}
C_{A11}=&-\frac{i(2\pi)^{6}}{N}\sum^{\infty}_{\boldsymbol{n}_{1}=-\infty}\iint\text{d}\boldsymbol{I}_{2}\,\text{d}\boldsymbol{I}_{3}\left[A_{\boldsymbol{n}_{1}\boldsymbol{0}}(\boldsymbol{I}_{1},\boldsymbol{I}_{2})\right]^{*}\nonumber\\
&\hspace{3.5cm}\times\boldsymbol{n}_{1}\cdot\frac{\partial}{\partial \boldsymbol{I}_{1}}\left\{\left[\frac{A_{\boldsymbol{n}_{1}\boldsymbol{0}}(\boldsymbol{I}_{1},\boldsymbol{I}_{3})}{-\boldsymbol{n}_{1}\cdot\boldsymbol{\Omega}_{1}+i\epsilon}\boldsymbol{n}\cdot\frac{\partial}{\partial \boldsymbol{I}_{1}}\right]f(\boldsymbol{I}_{1})\,f(\boldsymbol{I}_{2})\,f(\boldsymbol{I}_{3})\right\},\label{eq.CA11}
\end{align}
where equation \eqref{Eq.conjugate_A} is also employed to obtain the conjugate expression for the Fourier coefficients. Similarly, employing equations \eqref{Eq.inv_Fourier_phi}, \eqref{Eq.conjugate_A}, \eqref{Eq_delta_fuc}, and \eqref{eq:gA_expression}, the collision term $C_{A12}$, equation \eqref{Eq.collision_A12}, reduces to
\begin{align}
C_{A12}=&-\frac{i(2\pi)^{6}}{N}\sum^{\infty}_{\boldsymbol{n}_{1}=-\infty}\iint\text{d}\boldsymbol{I}_{2}\,\text{d}\boldsymbol{I}_{3}\left\{\boldsymbol{n}_{1}\cdot\frac{\partial}{\partial \boldsymbol{I}_{1}}\left[A_{\boldsymbol{n}_{1}\boldsymbol{0}}(\boldsymbol{I}_{1},\boldsymbol{I}_{2})\right]^{*}\right\}\nonumber\\
&\hspace{3cm}\times\left[\frac{A_{\boldsymbol{n}_{1}\boldsymbol{0}}(\boldsymbol{I}_{1},\boldsymbol{I}_{3})}{-\boldsymbol{n}_{1}\cdot\boldsymbol{\Omega}_{1}+i\epsilon}\boldsymbol{n}_{1}\cdot\frac{\partial}{\partial \boldsymbol{I}_{1}}\right]f(\boldsymbol{I}_{1})\,f(\boldsymbol{I}_{2})\,f(\boldsymbol{I}_{3}).\label{eq.CA12}
\end{align}
Combining equations \eqref{eq.CA11} and \eqref{eq.CA12}, one obtains
\begin{align}
&C_{A1}=-\frac{\pi(8\pi^{3})^{2}}{N}\sum^{\infty}_{\boldsymbol{n}_{1}=-\infty}\boldsymbol{n}_{1}\cdot\frac{\partial}{\partial \boldsymbol{I}_{1}}\left(\left|\int\text{d}\boldsymbol{I}_{2}A_{\boldsymbol{n}_{1}\boldsymbol{0}}(\boldsymbol{I}_{1},\boldsymbol{I}_{2})f(\boldsymbol{I}_{2},t)\right|^{2}\delta(\boldsymbol{n}_{1}\cdot\boldsymbol{\Omega}_{1})\,\boldsymbol{n}_{1}\cdot\frac{\partial f(\boldsymbol{I}_{1},t)}{\partial \boldsymbol{I}_{1}}\right).\label{Eq.collision_A1_gLandau_orb_result}
\end{align}
On the one hand, the correlation function $g_{A2}$ does not depend on the angle vector $\boldsymbol{\omega}_{1}$. Hence, the collision term $C_{A2}$ (due to the correlation function $g_{A2}$) reduces to $C_{A21}$. Lastly, after employing equations \eqref{Eq.inv_Fourier_phi}, \eqref{Eq_delta_fuc}, and \eqref{eq:gA_expression}, the collision term $C_{A21}$ vanishes as follows
\begin{subequations}
\begin{align}
C_{A2}&=C_{A21},\\
&=\frac{i(8\pi^{3})^{2}}{N}\sum^{\infty}_{\boldsymbol{n}=-\infty}\sum^{\infty}_{\boldsymbol{n}_{1}=-\infty}\sum^{\infty}_{\boldsymbol{n}_{2}=-\infty}\iint\text{d}\boldsymbol{I}_{2}\,\text{d}\boldsymbol{I}_{3}\,A_{\boldsymbol{n}_{1}\boldsymbol{n}_{2}}(\boldsymbol{I}_{1},\boldsymbol{I}_{2})\,\delta_{\boldsymbol{n},\boldsymbol{n}_{2}}\,\delta_{\boldsymbol{n}_{1},\boldsymbol{0}}\nonumber\\
&\hspace{5.4cm}\times\boldsymbol{n}_{1}\cdot\frac{\partial}{\partial \boldsymbol{I}_{1}}\left(\frac{A_{\boldsymbol{n}\boldsymbol{0}}(\boldsymbol{I}_{2},\boldsymbol{I}_{3})f(\boldsymbol{I}_{3},t)}{-\boldsymbol{n}\cdot\boldsymbol{\Omega}_{2}+i\epsilon}\,\boldsymbol{n}\cdot\frac{\partial f(\boldsymbol{I}_{1},t)\,f(\boldsymbol{I}_{2},t)}{\partial \boldsymbol{I}_{2}}\right),\\
&=0.
\end{align}
\end{subequations}
As a result, $C_{A}=C_{A1}$. Hence, we reach the expression of the collision term in equation \eqref{Eq.C_A_final}.

\section{The mathematical structure of the double-Fourier coefficients}\label{sec:fourier_coeff}

Thanks to the linearity of the action angles in the Hamiltonian equation \eqref{Eq.Hamilton_eqns_Iw}, the following mathematical operations have relatively clear physical meanings in the derivation of the orbit-averaged g-Landau equation 
\begin{subequations}
	\begin{align}
	\int \text{d}t &<=> \delta(\boldsymbol{n}\cdot\boldsymbol{\Omega}_{1}),\quad \text{or}\quad\delta(\boldsymbol{n}_{1}\cdot\boldsymbol{\Omega}_{1}-\boldsymbol{n}_{2}\cdot\boldsymbol{\Omega}_{2}),\hspace{0.5cm}\text{(Resonant condition)} \\
	\frac{\partial}{\partial \boldsymbol{\omega}}&<=> \boldsymbol{n}. \hspace{6cm} \text{(Orbital revolution)}
	\end{align}
\end{subequations}
The resonant condition shows the perturbation process due to the two-body or many-body relaxations, and the angular change measures the number of orbital revolutions. However, the double-Fourier coefficients $\{A_{\boldsymbol{nm}}(\boldsymbol{I}_{1},\boldsymbol{I}_{2})\}$ are more complicated to understand than the rest of the quantities. Because of its definition, equation \eqref{Eq.Fourier_phi}, the coefficients readily holds the conditions
\begin{subequations}
\begin{align}
&A_{\boldsymbol{nm}}(\boldsymbol{I}_{1},\boldsymbol{I}_{2})=\left[A_{\boldsymbol{mn}}(\boldsymbol{I}_{2},\boldsymbol{I}_{1})\right]^{*},\qquad (\text{self-adjoint})\label{Eq_self_adjo}\\
&A_{-\boldsymbol{n}-\boldsymbol{m}}(\boldsymbol{I}_{1},\boldsymbol{I}_{2})=\left[A_{\boldsymbol{nm}}(\boldsymbol{I}_{1},\boldsymbol{I}_{2})\right]^{*}. \label{Eq_conju}
\end{align}
\end{subequations}
A special case is the coefficients for the MF potential
\begin{subequations}
	\begin{align}
	&\Phi(\boldsymbol{I}_{1},\boldsymbol{\omega})=8\pi^{3}\sum_{\boldsymbol{n}_{1}=\boldsymbol{-\infty}}^{\boldsymbol{\infty}}\int \text{d}\boldsymbol{I}_{2}\, e^{i\boldsymbol{n}_{1}\cdot\boldsymbol{\omega}_{1}} A_{\boldsymbol{n}_{1}\boldsymbol{0}}(\boldsymbol{I}_{1},\boldsymbol{I}_{2})\,f(\boldsymbol{I}_{2},t),\\
	&A_{\boldsymbol{n}_{1}\boldsymbol{0}}(\boldsymbol{I}_{1},\boldsymbol{I}_{2})=\frac{1}{8\pi^{3}}\int\text{d}\boldsymbol{\omega}_{1}\,\bar{\phi}_{12}\,e^{-i\boldsymbol{n}_{1}\cdot\boldsymbol{\omega}_{1}},
	\end{align}
\end{subequations}
where the second subscript `$\boldsymbol{0}$' in the coefficients stands for the effect of the MF potential, and $\bar{\phi}_{12}$ is the orbit-averaged pairwise potential
\begin{align}
\bar{\phi}_{12}=\frac{1}{8\pi^{2}}\int\text{d}\boldsymbol{\omega}_{1}\, \phi_{12}.
\end{align}

To further understand the mathematical structure of the Fourier coefficients, we approximate them and the orbit-averaged g-Landau equation. We consider the following two special cases. (i) The discreteness fluctuation $\tilde{\phi}$ vanishes (Appendix \ref{sec.fine_DF}). (ii)  Either of or both the action vector and angle variables do not change during the relaxation process (Appendix \ref{sec:conserv_action_integrals}). These treatments are not proper for actual applications. However, they can provide approximates of the orbit-averaged g-Landau kinetic equation. Lastly, Appendix \ref{Appendix:covention_Landau} shows an analogous relationship between the g-Landau equation and the FP equation with a homogeneous background.

\subsection{`Discreteness fluctuation' and double Fourier coefficients}\label{sec.fine_DF}
The present appendix clarifies the relationship between the discreteness fluctuation $\tilde{\phi}$ and the Fourier coefficients $\{A_{\boldsymbol{nm}}(\boldsymbol{I}_{1},\boldsymbol{I}_{2})\}$ by assuming that the discrete fluctuation does \emph{not} occur.  The orbit-averaged squares of MF- and pairwise Newtonian potentials may be expressed in terms of $\{A_{\boldsymbol{nm}}(\boldsymbol{I}_{1},\boldsymbol{I}_{2})\}$ as
\begin{subequations}
	\begin{align}
	&\frac{1}{8\pi^{3}}\int\text{d}\boldsymbol{\omega}_{1}\left|\Phi(1,t)\right|^{2}=8\pi^{3}\sum_{\boldsymbol{n}_{1}=-\infty}^{\infty}\left|\int\text{d}\boldsymbol{I}_{2}\,A_{\boldsymbol{n}_{1}0}(\boldsymbol{I}_{1},\boldsymbol{I}_{2})\right|^{2},\\
	&\frac{1}{8\pi^{3}}\int\text{d}\boldsymbol{\omega}_{1}\left|\phi_{12}\right|^{2}=\sum_{\boldsymbol{n}_{1}=-\infty}^{\infty}\sum_{\boldsymbol{n}_{2}=-\infty}^{\infty}\left|A_{\boldsymbol{n}_{1}\boldsymbol{n}_{2}}(\boldsymbol{I}_{1},\boldsymbol{I}_{2})\right|^{2}.
	\end{align}\label{Eq.orb_ave_Potential}
\end{subequations}
The mean square of the discrete fluctuation in the pairwise Newtonian potential reads
\begin{align}
\left<\tilde{\phi}^{2}_{12}\right>&=\frac{1}{N}\left(\int\text{d}2\, f(2, t)\left|\phi_{12} \right|^{2}-\frac{1}{N}\left|\Phi(1,t)\right|^{2}\right),\label{Eq.sqrt_fluctu}
\end{align}
where 
\begin{align}
\tilde{\phi}_{12}=\phi_{12}-\dfrac{\Phi(1,t)}{N}.
\end{align}
Employing equations \eqref{Eq.orb_ave_Potential} and \eqref{Eq.sqrt_fluctu},  we can find the orbit-averaged expression of equation \eqref{Eq.sqrt_fluctu} in terms of the coefficients $\{A_{\boldsymbol{nm}}(\boldsymbol{I}_{1},\boldsymbol{I}_{2})\}$  as follows
\begin{align}
\frac{1}{8\pi^{3}}\int\text{d}\boldsymbol{\omega}_{1}\left<\tilde{\phi}^{2}_{12}\right>&=(8\pi^{3})\sum^{\boldsymbol{\infty}}_{\boldsymbol{n}=\boldsymbol{-\infty}}\sum^{\boldsymbol{\infty}}_{\boldsymbol{n}_{1}=-\boldsymbol{\boldsymbol{\infty}}\&\boldsymbol{n}_{1}\neq\boldsymbol{0}}\int\text{d}\boldsymbol{I}_{2}\left|A_{\boldsymbol{n}\boldsymbol{n}_{1}}(\boldsymbol{I}_{1},\boldsymbol{I}_{2})\right|^{2}f(\boldsymbol{I}_{2},t)\nonumber\\
&+\frac{(8\pi^{3})^{2}}{2N}\sum^{\boldsymbol{\infty}}_{\boldsymbol{n}=\boldsymbol{-\infty}}\int\text{d}\boldsymbol{I}_{2}\int\text{d}\boldsymbol{I}_{3}\left|A_{\boldsymbol{n}\boldsymbol{0}}(\boldsymbol{I}_{1},\boldsymbol{I}_{2})-A_{\boldsymbol{n}\boldsymbol{0}}(\boldsymbol{I}_{1},\boldsymbol{I}_{3})\right|^{2}f(\boldsymbol{I}_{2},t)f(\boldsymbol{I}_3,t).\label{Eq.phi_disp}
\end{align}
Hence, the orbit-averaged discrete fluctuation can reproduce the  essential mathematical structures of the collision terms in the orbit-averaged g-Landau equation \eqref{Eq.orb_gLandau}, except for the resonant condition and the number of orbital revolutions. 

As pointed out in Section \ref{sec:import_discrete}, star 3 represents the effect of many-body relaxation, i.e., the discreteness stochastic acceleration. Our concern is how the Fourier coefficients are reduced after we artificially assume that the discreteness stochastic acceleration does not happen. This assumption corresponds with the operation that the second line in equation \eqref{Eq.phi_disp} vanishes. Accordingly,  we have the following equality for any $\boldsymbol{I}_{2}$ and $ \boldsymbol{I}_{3}$
\begin{align}
A_{\boldsymbol{n} \boldsymbol{0}}(\boldsymbol{I}_{1},\boldsymbol{I}_{2})=A_{\boldsymbol{n}\boldsymbol{0}}(\boldsymbol{I}_{1},\boldsymbol{I}_{3})\equiv A_{\boldsymbol{n}\boldsymbol{0}}(\boldsymbol{I}_{1}). \label{Eq.non_mf_cond}
\end{align}
Equation \eqref{Eq.non_mf_cond} infers that the MF potential reduces to the sum of orbit-averaged pairwise potentials as follows
\begin{subequations}
	\begin{align}
	&\Phi(\boldsymbol{I}_{1},\boldsymbol{\omega}_{1})=N\sum_{\boldsymbol{n}_{1}=-\infty}^{\infty} e^{i\boldsymbol{n}_{1}\cdot\boldsymbol{\omega}_{1}} A_{\boldsymbol{n}_{1}\boldsymbol{0}}(\boldsymbol{I}_{1})=N\bar{\phi}_{12}(\boldsymbol{I}_{1},\boldsymbol{\omega}_{1}),\\
	&A_{\boldsymbol{n}_{1}0}(\boldsymbol{I}_{1})=\frac{1}{8\pi^{3}}\int\text{d}\boldsymbol{\omega}_{1}\, \bar{\phi}_{12}(\boldsymbol{I}_{1},\boldsymbol{\omega}_{1})\, e^{-i\boldsymbol{n}_{1}\cdot\boldsymbol{\omega}_{1}}.
	\end{align}
\end{subequations}

\subsection{Conservation of action-angle variables} \label{sec:conserv_action_integrals}

We approximate the double-Fourier coefficients and the g-Landau equation into simpler forms by artificially assuming the conservation laws of  the total action vector (Appendix \ref{Appendix:conserv_integral}) and the total action angle (Appendix \ref{Appendix:conserv_angle}). 

\subsubsection{Conservation of the total integral vector}\label{Appendix:conserv_integral}

We discuss the collision term $C\left[f(\boldsymbol{I},t)\right]$, equation \eqref{Eq.collision_gLandau_orb}, in the g-Landau kinetic equation in action-angle variables. First, assume that the conservation of the integral vector holds for the equation
\begin{align}
    \frac{\partial}{\partial t}\left(\int \text{d}1\,\boldsymbol{I}_{1}\,f(\boldsymbol{I}_{1},t)\right)=\int \text{d}1\,\boldsymbol{I}_{1}\,C\left[f(\boldsymbol{I}_{1},t)\right]=0.\label{Eq.consv_I}
\end{align}
Because of the first equation of the BBGKY hierarchy, equation \eqref{Eq.1stBBGKY_gLandau_orb}, the total integral vector is expressed as
\begin{align}
\int \text{d}1\,\boldsymbol{I}_{1}\,C\left[f(\boldsymbol{I}_{1},t)\right]&=\iint\text{d}1\,\text{d}2\,\boldsymbol{I}_{1}\left(\frac{\partial \phi_{12}}{\partial \boldsymbol{\omega}_{1}}\cdot\frac{\partial g(1,2,t)}{\partial \boldsymbol{I}_{1}}-\frac{\partial \phi_{12}}{\partial \boldsymbol{I}_{1}}\cdot\frac{\partial g(1,2,t)}{\partial \boldsymbol{\omega}_{1}}\right),\nonumber\\
&=-\iint\text{d}1\,\text{d}2\,\frac{\partial \phi_{12}}{\partial \boldsymbol{\omega}_{1}}\,g(1,2,t).\label{Eq.tot_I1}
\end{align}
By employing the definition of the double Fourier coefficients, equation \eqref{Eq.inv_Fourier_phi}, and their properties, equations \eqref{Eq_self_adjo} and    \eqref{Eq_conju}, we can rewrite equation \eqref{Eq.tot_I1} as follows
\begin{equation}
\int \text{d}1\boldsymbol{I}_{1}C\left[f(\boldsymbol{I}_{1},t)\right]=-\frac{i}{2}\iint\text{d}1\,\text{d}2 \sum_{\boldsymbol{n}_{1}=-\infty}^{\infty}\sum_{\boldsymbol{n}_{2}=-\infty}^{\infty} g(1,2,t)\, (\boldsymbol{n}_{1}-\boldsymbol{n}_{2})\, A_{\boldsymbol{n}_{1}\boldsymbol{n}_{2}}(\boldsymbol{I}_{1},\boldsymbol{I}_{2})\, e^{i(\boldsymbol{n}_{1}\cdot\boldsymbol{\omega}_{1}-\boldsymbol{n}_{1}\cdot\boldsymbol{\omega}_{2})}.\label{Eq.tot_I1_2}
\end{equation}
Hence, the conservation of the total integral vector, equation \eqref{Eq.consv_I}, can hold for any action-angle variables if the following condition is valid
\begin{align}
\boldsymbol{n}_{1}=\boldsymbol{n}_{2}.\label{Eq.cond_conserv_integrals}
\end{align}
This relationship imposes a  condition on the Fourier coefficients, equation \eqref{Eq.inv_Fourier_phi}, that the potential $\phi_{12}$ is a function of $(\boldsymbol{\omega}_{1}-\boldsymbol{\omega}_{2})$. Then, one may Fourier-transform the potential as follows
\begin{subequations}
	\begin{align}
	&\phi_{1i}=\sum_{\boldsymbol{n}_{1}=-\infty}^{\infty}A_{\boldsymbol{n}_{1}\boldsymbol{n}_{1}}(\boldsymbol{I}_{1},\boldsymbol{I}_{i})\, e^{i\boldsymbol{n}_{1}\cdot(\boldsymbol{\omega}_{1}-\boldsymbol{\omega}_{i})},\qquad (i=2\text{ or } 3)\\
	&A_{\boldsymbol{n}_{1}\boldsymbol{n}_{1}}(\boldsymbol{I}_{1},\boldsymbol{I}_{i})=\frac{1}{8\pi^{3}}\int\text{d}\boldsymbol{\omega}_{1i}\, \phi_{1i}(\omega_{1i},\boldsymbol{I}_{1})\, e^{-i\boldsymbol{n}_{1}\cdot\boldsymbol{\omega}_{1i}}.
	\end{align}
\end{subequations} 
Lastly, we simplify the orbit-averaged g-Landau kinetic equation \eqref{Eq.orb_gLandau} using the conditions above. Assume that the discreteness stochastic acceleration does not occur (equation \eqref{Eq.non_mf_cond}) and the conservation of the total integral vector holds (equation \eqref{Eq.cond_conserv_integrals}). Then, the g-Landau equation reduces to
\begin{align}
\partial_{t}f(\boldsymbol{I}_{1},t)&=\pi(8\pi^{3})\sum^{\infty}_{\boldsymbol{n}=-\infty}\int\text{d}\boldsymbol{I}_{2}\, \boldsymbol{n}\cdot\frac{\partial}{\partial \boldsymbol{I}_{1}}\nonumber\\
&\hspace{1cm}\times\left\{\left|A_{\boldsymbol{n}\boldsymbol{n}}(\boldsymbol{I}_{1},\boldsymbol{I}_{2})\right|^{2}\, \delta(\boldsymbol{n}\cdot[\boldsymbol{\Omega}_{1}-\boldsymbol{\Omega}_{2}])\, \boldsymbol{n}\cdot\left(\frac{\partial}{\partial \boldsymbol{I}_{1}}-\cdot\frac{\partial}{\partial \boldsymbol{I}_{2}}\right)f(\boldsymbol{I}_{1},t)f(\boldsymbol{I}_{2},t)\right\}.
\end{align}

\subsubsection{Conservation of action angles}\label{Appendix:conserv_angle}

Similarly to the conservation of the total integral vector discussed in Appendix \ref{Appendix:conserv_integral}, we consider the conservation of the total action angles as follows
\begin{align}
    \frac{\partial}{\partial t}\left(\int \text{d}1\,\boldsymbol{\omega}_{1}\,f(\boldsymbol{I}_{1},t)\right)=\int \text{d}1\,\boldsymbol{\omega}_{1}\,C\left[f(\boldsymbol{I}_{1},t)\right]=0.
\end{align}
Using equation \eqref{Eq.inv_Fourier_phi}, the total action angle reduces to
\begin{subequations}
	\begin{align}
	\int \text{d}1\,\boldsymbol{\omega}_{1}\,C\left[f(\boldsymbol{I}_{1},t)\right]&=\iint\text{d}1\,\text{d}2\,\frac{\partial \phi_{12}}{\partial \boldsymbol{I}_{1}}\,g(1,2,t)=\iint\text{d}1\,\text{d}2\,\frac{\partial \phi_{12}}{\partial \boldsymbol{I}_{2}}\,g(1,2,t),\label{Eq.tot_omega1}\\
	&=\iint\text{d}1\,\text{d}2 \sum_{\boldsymbol{n}_{1}=-\infty}^{\infty}\sum_{\boldsymbol{n}_{2}=-\infty}^{\infty}\, g(1,2,t)\,\frac{\partial}{\partial \boldsymbol{I}_{1}}A_{\boldsymbol{n}_{1}\boldsymbol{n}_{2}}(\boldsymbol{I}_{1},\boldsymbol{I}_{2})\, e^{i(\boldsymbol{n}_{1}\cdot\boldsymbol{\omega}_{1}-\boldsymbol{n}_{1}\cdot\boldsymbol{\omega}_{2})}.\label{Eq.tot_omega1_2}
	\end{align}
\end{subequations}
If we impose the following strong condition
\begin{align}
\frac{\partial}{\partial \boldsymbol{I}_{1}}A_{\boldsymbol{n}_{1}\boldsymbol{n}_{2}}(\boldsymbol{I}_{1},\boldsymbol{I}_{2})=0=\frac{\partial}{\partial \boldsymbol{I}_{2}}A_{\boldsymbol{n}_{1}\boldsymbol{n}_{2}}(\boldsymbol{I}_{1},\boldsymbol{I}_{2})\qquad\quad=>\qquad A_{\boldsymbol{n}_{1}\boldsymbol{n}_{2}}(\boldsymbol{I}_{1},\boldsymbol{I}_{2})\equiv A_{\boldsymbol{n}_{1}\boldsymbol{n}_{2}},\label{Eq.cond_conserv_angles}
\end{align}
the corresponding pairwise and MF potentials are written as
\begin{subequations}
	\begin{align}
	&\phi_{1i}=\sum_{\boldsymbol{n}_{1}=-\infty}^{\infty}\sum_{\boldsymbol{n}_{i}=-\infty}^{\infty}A_{\boldsymbol{n}_{1}\boldsymbol{n}_{i}}\, e^{i(\boldsymbol{n}_{1}\cdot\boldsymbol{\omega}_{1}-\boldsymbol{n}_{i}\cdot\boldsymbol{\omega}_{i})},\qquad (i=2\text{ or } 3)\\
	&A_{\boldsymbol{n}_{1}\boldsymbol{n}_{i}}=\frac{1}{(8\pi^{3})^{2}}\iint\text{d}\boldsymbol{\omega}_{1}\, \text{d}\boldsymbol{\omega}_{i}\, \phi_{1i}(1,i)\, e^{-i(\boldsymbol{n}_{1}\cdot\boldsymbol{\omega}_{1}-\boldsymbol{n}_{i}\cdot\boldsymbol{\omega}_{i})},\\
	&\Phi(\boldsymbol{\omega}_{1})=N\sum_{\boldsymbol{n}_{1}=-\infty}^{\infty} e^{i\boldsymbol{n}_{1}\cdot\boldsymbol{\omega}_{1}} A_{\boldsymbol{n}_{1}\boldsymbol{0}}=N\bar{\phi}_{12}(\boldsymbol{\omega}_{1}).
	\end{align}
\end{subequations}
Then, the g-Landau equation equation \eqref{Eq.orb_gLandau} reduces to
\begin{align}
\partial_{t}f(\boldsymbol{I}_{1},t)=&\pi(8\pi^{3})\sum^{\infty}_{\boldsymbol{n}=-\infty}\sum^{\infty}_{\boldsymbol{n}_{1}=-\infty\&\boldsymbol{n}_{1}\neq\boldsymbol{0}}\left|A_{\boldsymbol{n}\boldsymbol{n}_{1}}\right|^{2}\, \int\text{d}\boldsymbol{I}_{2}\, \boldsymbol{n}\cdot\frac{\partial}{\partial \boldsymbol{I}_{1}}\\
&\hspace{3cm}\times\left\{\delta(\boldsymbol{n}\cdot\boldsymbol{\Omega}_{1}-\boldsymbol{n}_{1}\cdot\boldsymbol{\Omega}_{2})\left(\boldsymbol{n}\cdot\frac{\partial}{\partial \boldsymbol{I}_{1}}-\boldsymbol{n}_{1}\cdot\frac{\partial}{\partial \boldsymbol{I}_{2}}\right)f(\boldsymbol{I}_{1},t)\,f(\boldsymbol{I}_{2},t)\right\}.
\end{align}

\subsection{Deriving the local FP kinetic equation for stars under a homogeneous background approximation}\label{Appendix:covention_Landau}

One can reduce the orbit-averaged g-Landau equation  \eqref{Eq.orb_gLandau} to the local FP equation with a homogeneous background if holding the conservation of both the integral vector and the angle vector.  We assume the conservation laws, equations \eqref{Eq.cond_conserv_integrals} and \eqref{Eq.cond_conserv_angles} to  understand this statement. Then, the laws reduce the pairwise and MF potentials to 
\begin{subequations}
	\begin{align}
	&\phi_{1i}=\sum_{\boldsymbol{n}_{1}=-\infty}^{\infty}A_{\boldsymbol{n}_{1}\boldsymbol{n}_{1}}\, e^{i\boldsymbol{n}_{1}\cdot(\boldsymbol{\omega}_{1}-\boldsymbol{\omega}_{i})},\qquad (i=2\text{ or } 3)\\
	&A_{\boldsymbol{n}_{1}\boldsymbol{n}_{i}}=\frac{1}{(8\pi^{3})}\int\text{d}\boldsymbol{\omega}_{1i}\, \phi_{1i}(1,i)\, e^{-i\boldsymbol{n}_{1}\cdot(\boldsymbol{\omega}_{1}-\boldsymbol{\omega}_{i})},\\
	&\Phi(\boldsymbol{\omega}_{1})=0.
	\end{align}
\end{subequations}
Accordingly, we find a one-to-one relationship between the relative distance $r_{12}$ and relative angle $\omega_{12}$ between two stars;
\begin{align}
\phi_{12}=-\frac{Gm}{\left|\boldsymbol{r}_{1}(\boldsymbol{\omega}_{1})-\boldsymbol{r}_{2}(\boldsymbol{\omega}_{2})\right|}=-\frac{Gm}{\left|\boldsymbol{r}_{12} (\boldsymbol{\omega}_{12})\right|}.
\end{align}
This relationship means that one needs only the modulus of the relative displacement vector $r_{12}(=\left|\boldsymbol{r}_{12}\right|)$ and the corresponding Fourier variable, the `wavenumber', $\omega_{12}=\left|\boldsymbol{\omega}_{12}\right|$. Also, because of the absence of the MF potential, the one-body Hamiltonian $h$, equation \eqref{Eq.one_body_Hamilton}, reduces to
\begin{align}
h=\frac{\boldsymbol{p}_{1}^{2}}{2m}.
\end{align}
Hence, the corresponding actions and angles may be analogically labeled to describe the test star's rectilinear motion as follows
\begin{subequations}
	\begin{align}
	&\boldsymbol{I}_{1}=\boldsymbol{p}_{1}^{2}=\text{const.},\\
	&\boldsymbol{\omega}_{1}=\boldsymbol{v}_{1},\\
	&\boldsymbol{\omega}_{1}=\boldsymbol{r}_{1}.
	\end{align}
\end{subequations}
As a result, one obtains the following equation from the g-Landau equation  \eqref{Eq.orb_gLandau}
	\begin{align}
	\partial_{t}f(\boldsymbol{p}_{1},t)=&\pi(8\pi^{3})\sum^{\infty}_{\boldsymbol{n}=-\infty}\left|A_{\boldsymbol{n}\boldsymbol{n}}\right|^{2}\int\, \text{d}\boldsymbol{p}_{2}\,\boldsymbol{n}\cdot\frac{\partial}{\partial \boldsymbol{p}_{1}}\nonumber\\
	&\hspace{3cm}\times\left\{\delta(\boldsymbol{n}\cdot\left[\boldsymbol{v}_{1}-\cdot\boldsymbol{v}_{2}\right])\left(\boldsymbol{n}\cdot\left[\frac{\partial}{\partial \boldsymbol{p}_{1}}-\frac{\partial}{\partial \boldsymbol{p}_{2}}\right]\right)f(\boldsymbol{p}_{1},t)\,f(\boldsymbol{p}_{2},t)\right\}.\label{Eq.g-Landau_homo_disc}
	\end{align}
In physical spaces, the pairwise Newtonian force dominates stars' motions on scales smaller than the ``encounter radius'' \citep{Ogorodnikov_1965}. To count the effect of stars' non-periodicity on small scales, one must employ the continuous Fourier transform in place of the discreteness Fourier-series expansion of the potential. With the continuous transform, equation \eqref{Eq.g-Landau_homo_disc} analogically reduces to the FP equation with a homogeneous background
\begin{align}
	\partial_{t}f(\boldsymbol{p}_{1},t)\approx&\pi(8\pi^{3})\int\text{d}\boldsymbol{n}\left|A_{\boldsymbol{n}\boldsymbol{n}}\right|^{2}\int\text{d}\boldsymbol{p}_{2}\, \boldsymbol{n}\cdot\frac{\partial}{\partial \boldsymbol{p}_{1}}\nonumber\\
	&\hspace{3cm}\times\left\{\delta(\boldsymbol{n}\cdot\left[\boldsymbol{v}_{1}-\cdot\boldsymbol{v}_{2}\right])\left(\boldsymbol{n}\cdot\left[\frac{\partial}{\partial \boldsymbol{p}_{1}}-\frac{\partial}{\partial \boldsymbol{p}_{2}}\right]\right)f(\boldsymbol{p}_{1},t)\,f(\boldsymbol{p}_{2},t)\right\}.
\end{align}
A further explicit form of the FP equation can be found in \citep[e.g.,][]{Kandrup_1981,Chavanis_2013a}.

\end{appendices}

\bibliographystyle{raa}
\bibliography{science}


\label{lastpage}

\end{document}